\documentclass[12pt]{article}
\usepackage[utf8,ansinew]{inputenc}

\pdfoutput=1

\usepackage{jheppub}
\usepackage{appendix}
\usepackage{color}
\usepackage{epsfig, palatino}
\usepackage{epic}
\usepackage{mathrsfs}
\usepackage{ae} 
\usepackage[T1]{fontenc}
\usepackage{amsmath}
\usepackage{physics}
\usepackage{bbm}
\usepackage{amssymb}
\usepackage{bm}
\usepackage{graphicx}
\usepackage{tensor}
\usepackage{ulem}
\usepackage{mathtools}
\usepackage[skip=10pt plus1pt]{parskip}
\usepackage{empheq}
\usepackage[most]{tcolorbox}
\usepackage[export]{adjustbox}

\newtcbox{\mymath}[1][]{
    nobeforeafter, math upper, tcbox raise base,
    enhanced, colframe=blue!30!black,
    boxrule=1pt,
    #1}

\def\be#1\ee{\begin{align}#1\end{align}}

\definecolor{darkblue}{cmyk}{0.9,0.9,0,0}
\definecolor{darkgreen}{cmyk}{0.9,0,0.9,0}
\definecolor{blueblue}{cmyk}{0.73,0.28,0,0.5}
\definecolor{lightblue}{RGB}{55,171,200}
\definecolor{grey}{gray}{0.55}
\definecolor{pink}{cmyk}{0., 0.9859943977591037, 0.3571428571428571, 0.16000000000000003}
\definecolor{lightpink}{cmyk}{0., 0.5, 0.5, 0.}
\definecolor{lightgreen}{cmyk}{0.24175824175824182, 0., 0.9615384615384616, 0.28627450980392155}

\usepackage{hyperref}
\usepackage{wasysym}
\usepackage{varioref}
\usepackage{makeidx}
\usepackage[english]{babel}
\usepackage{simplewick}
\usepackage{array}
\usepackage{multirow}
\usepackage{tabularx}

\usepackage{caption}
\usepackage{subcaption}

\def\({\left(}
\def\){\right)}
\def\[{\left[}
\def\]{\right]}
\def\<{\langle}
\def\>{\rangle}

\def\d{\mathrm{d}}

\newcommand{\beq}{\begin{equation}}
	\newcommand{\eeq}{\end{equation}}
\newcommand{\beqq}{\begin{equation*}}
	\newcommand{\eeqq}{\end{equation*}}
\newcommand\beqa{\begin{eqnarray}}
	\newcommand\eeqa{\end{eqnarray}}

\usepackage{tikz}
\usetikzlibrary { decorations.pathmorphing, decorations.pathreplacing, decorations.shapes, }
\usetikzlibrary{shapes.misc}

\tikzset{cross/.style={cross out, draw=black, minimum size=2*(#1-\pgflinewidth), inner sep=0pt, outer sep=0pt},
cross/.default={1pt}}

\title{The Born regime of gravitational amplitudes}
\author[a]{Miguel Correia}
\author[b]{and Giulia Isabella}
\affiliation[a]{Department of Physics, McGill University, 3600 Rue University, Montr\'eal, H3A 2T8, QC Canada}
\affiliation[b]{Mani L. Bhaumik Institute for Theoretical Physics, Department of Physics and Astronomy, University of California Los Angeles, Los Angeles, CA 90095, USA}

\emailAdd{miguel.ribeirocorreia@mcgill.ca, giuliaisabella@physics.ucla.edu}

 \abstract{
We study the $2 \to 2$ scattering in the regime where the wavelength of the scattered objects is comparable to their distance but is much larger than any Compton wavelength in the quantum field theory. 
We observe that in this regime - which differs from the eikonal - the Feynman diagram expansion takes the form of a geometric series, akin to the Born series of quantum mechanics.
Conversely, we can define the Feynman diagram expansion as the Born series of a relativistic effective-one-body (EOB) Schr{\"o}dinger equation. For a gravitational theory in this regime we observe that the EOB Schr{\"o}dinger equation reduces to the Regge-Wheeler or Teukolsky wave equations.
We make use of this understanding to study the tree-level Compton scattering off a Kerr black hole. We compute the scalar and photon Compton amplitude up to $O(a^{30})$ in the black hole spin $a$ and propose an all-order expression. Remarkably, we find that boundary terms, which are typically neglected, give non-zero contact pieces necessary for restoring crossing symmetry and gauge invariance of the Kerr-Compton amplitude.

}

\begin{document}

\maketitle

\section{Introduction}
The question of recovering classical physics from Quantum Field Theory (QFT) is undoubtedly an old one, but it has undergone an intense revival in recent years driven by the detection of Gravitational Waves (GW) \cite{LIGOScientific:2016aoc,KAGRA:2020agh}. 
 In the  eikonal regime, which applies in the inspiral phase of black hole (BH) binaries, 
scattering amplitudes have allowed us to determine classical observables to impressive precision in the so-called Post-Minkowskian (PM) expansion \cite{Goldberger:2004jt,Rothstein:2014sra,Goldberger:2022rqf,Porto:2016pyg,Kalin:2020mvi,Mogull:2020sak,Cheung:2018wkq,Kosower:2018adc,Bern:2019crd,Bern:2020buy,Buonanno:2022pgc,Bjerrum-Bohr:2022blt,Bjerrum-Bohr:2021din,Bjerrum-Bohr:2014zsa,Bern:2021yeh,Buonanno:1998gg,Buonanno:2000ef,Kalin:2020fhe,Dlapa:2021npj,Driesse:2024xad,Bern:2024adl} (and references therein). A key feature of the eikonal regime is the exponentiation of the amplitude in angular momentum space in terms of a classical action known as the radial action \cite{FORD1959287,Berry_1972,Kol:2021jjc, Bellazzini:2022wzv,Bern:2021dqo}. Since it is from the radial action that classical observables can be determined, it is important to identify in the perturbative amplitude computation which terms contain genuine new physics, and which contributions are just iterations of lower order pieces of the radial action (typically known as superclassical terms) \cite{Amati:1987uf,Amati:1987wq,Amati:1990xe,Amati:1992zb,Kabat:1992tb,Bellazzini:2022wzv,Bern:2021dqo,Herrmann:2021tct,DiVecchia:2021bdo,Brandhuber:2021eyq,Bjerrum-Bohr:2021vuf,DiVecchia:2023frv}. 

On the other hand, the regime relevant for the ringdown phase, where GW fluctuations scatter off a  single black hole background \cite{Anninos:1993zj,Berti:2009kk,Vishveshwara:1970zz,Chandrasekhar:1975zza,PhysRevLett.27.1466,Kokkotas:1999bd,Konoplya:2011qq}, has been less explored in the context of scattering amplitudes. The setup of this problem in General Relativity (GR) is well understood, and it boils down to solving classical wave equations known as the Teukolsky equations \cite{1973ApJ...185..635T,1973ApJ...185..649P,1974ApJ...193..443T}. The resulting scattering phase shift is determined in a low-frequency expansion by employing black hole perturbation theory (BHPT) methods \cite{Mano:1996vt,Mano:1996gn,Mano:1996mf,Sasaki:2003xr,Bonelli:2022ten,Dodelson:2022yvn,Aminov:2023jve}. This information can then be translated using amplitude techniques to an effective field theory (EFT) description of black holes as point-particles that couple non-minimally to gravity. These non-minimal couplings include information about dissipation and tidal deformability of BHs, known as tidal Love numbers \cite{Goldberger:2005cd,Goldberger:2020fot,Goldberger:2020wbx,Ivanov:2022qqt,Saketh:2023bul,Ivanov:2024sds}, or sub-leading spin effects in the case of Kerr BHs \cite{Porto:2005ac,Levi:2015msa,Saketh:2022wap,Jakobsen:2023ndj,Ben-Shahar:2023djm,Arkani-Hamed:2019ymq,Guevara:2018wpp,Chung:2019duq,Aoude:2020onz,Chung:2018kqs,Arkani-Hamed:2017jhn,Bautista:2021wfy,Bautista:2022wjf,Bautista:2023sdf,Scheopner:2023rzp,Vines:2017hyw,Kosmopoulos:2021zoq,Chen:2021kxt,Aoude:2022trd,Bern:2022kto,Aoude:2022thd,FebresCordero:2022jts,Menezes:2022tcs,Bjerrum-Bohr:2023jau,Aoude:2023vdk,Bern:2023ity,Bjerrum-Bohr:2023iey,Chen:2023qzo,Luna:2023uwd,Guevara:2019fsj,Cangemi:2022bew,Cangemi:2022abk,Cangemi:2023ysz,Chia:2020yla,Cangemi:2023bpe}. Once these EFT  coefficients are determined, they can be used to predict absorption or spin effects in a different context, for example in the late inspiral stage before the merger.

Contrarily to the eikonal regime, this regime is characterized by the wavelength of the scattering objects being comparable to their distance, with QFT effects such as pair production still being absent. In the perturbative Feynman diagram computation it is still not quite yet understood which pieces survive, and whether a concept of iteration exists. In this work we address this question and shed some light on the general structure of scattering amplitudes in this regime.

We find that in this regime, which we call the Born regime, the Feynman diagram expansion reduces to a geometric series akin to the Born series of quantum mechanics. In the Born regime the object that is naturally iterated is an effective potential, which enjoys a larger analyticity domain than the scattering amplitude, among other useful properties. With the potential one can define an effective-one-body (EOB) relativistic Schr{\"o}dinger equation \cite{Todorov:1970gr}, which has been used to compute radiative corrections to the energy levels of bound states such as positronium or the Hydrogen atom \cite{Rizov:1975tr,Crater:1992xp,Jallouli:1996bu,Berestetskii:1982qgu}.\footnote{The EOB terminology is typically used in the GR context referring to the Buonanno-Damour method to handle the gravitational two-body problem \cite{Buonanno:1998gg}. However, we believe it is also a good terminology for Todorov's framework in QFT \cite{Todorov:1970gr}, which preceded that of Buonanno and Damour, and later shown to be equivalent to that of GR in the classical (eikonal) limit \cite{Fiziev:2000mh}.} In the Born regime of gravitational scattering, we observe that the EOB Schr{\"o}dinger equation reduces to a classical wave equation, such as the Regge-Wheeler \cite{PhysRev.108.1063} or Teukolsky  equations. This understanding establishes a direct link between the QFT scattering amplitude computation and the GR equations of motion, allowing us in some cases to completely circumvent the (arguably cumbersome) conversion of BHPT results into QFT language, and vice-versa. 

This means of course that the computation of gravitational amplitudes in this regime can be aided by making use of established techniques in quantum mechanics, which typically involve Fourier transforms and Born-type integrals. One such application that has garnered some attention lately is the Compton  scattering off a Kerr black hole. It was first noted in \cite{Arkani-Hamed:2019ymq} that the scattering off a Kerr black hole at $O(G)$ is reproduced in the eikonal regime by taking the large spin limit of a set of minimally coupled massive spinning particles \cite{Arkani-Hamed:2017jhn}, with the amplitude exhibiting exponentiation in the classical spin $a$. Subsequent work \cite{Guevara:2018wpp,Aoude:2020onz,Aoude:2022trd,Cangemi:2022bew,Cangemi:2022abk,Cangemi:2023ysz} has showed that moving away from the eikonal regime requires the presence of contact terms which lack an appropriate definition of minimal coupling. Those contact terms affect the massive-massive two-body problem by entering in loops and changing its long-range behavior, consequentially capturing deviations from the point-particle description. Current approaches to fix contact terms rely on writing down the most general EFT of higher-spin massive particles and matching the EFT coefficients with the results from BHPT. However, extracting the amplitude at $O(G)$ and  higher orders in the spin $a$ from the BHPT results involves taking the super-extremal limit of Kerr BHs which requires a non-unique analytic continuation in $a/{GM}$  \cite{Bautista:2021wfy,Bautista:2022wjf,Bautista:2023sdf}.
\footnote{Recent progress however has been made in \cite{Bautista:2023sdf}, where a near-far factorization of the BHPT result allows to circumvent the analytic continuation and still captures some contact terms.
}

Here we use the insight from the Born regime to address this problem. We write the down the equations of motion for a massless scalar and photon scattering off a Kerr BH (which precede the Teukolsky equation). In particular, we make use of the Kerr-Schild coordinate system which allows us to isolate a Lorentz covariant potential. A ``simple" Fourier transform of this potential lets us compute the Kerr-Compton amplitude at $O(G)$ to very high order in the BH spin $a$ and resum the answer.

This work is organised as follows. In Chapter \ref{sec:scales} we discuss the relevant scales in gravitational scattering and compare the eikonal and Born regimes. In Chapter \ref{sec:EOB} we review the effective-one-body framework in QFT following Todorov \cite{Todorov:1970gr} and show how it reduces to a classical wave description in the Born regime. In Chapter \ref{sec:2PM} we consider the gravitational scattering of two minimally coupled scalars in the Born regime and show how the Born series of the Regge-Wheeler equation is recovered at $O(G^2)$. In Chapter \ref{sec:Kerr} we present the aforementioned calculation of the scalar and photon Kerr-Compton amplitudes. In Chapter \ref{sec:conclusion} we conclude our work and discuss some future directions.


\section{Classical regimes: Eikonal vs Born}\label{sec:scales}

For the scattering of two wavepackets in the center of mass frame we identify the following kinematical length scales:

\begin{itemize}
    \item Wavelength of the scattered wavepackets, $\lambda = \frac{\hbar}{|\mathbf{p}|}$, where $|\mathbf{p}|$ is the center of mass momentum\,\footnote{In this work, we adopt the \textbf{bold} notation to refer to 3-vectors.}. This is typically the length scale where effects associated with the wave nature of the particle become important. For massive particles $\lambda = \lambda_\text{dB}$ is the quantum mechanical de Broglie wavelength, whereas if one of the particles is massless then $\lambda = \frac{1}{\omega}$ is the wavelength of the massless wave with frequency $\omega$.
     \item Impact parameter $b$ between the scattered wavepackets, which is related to the angular momentum by $J = b |\mathbf{p}|$. The impact parameter is Fourier conjugate to the transferred momentum $|\mathbf{q}|$. Generically, we expect $b \sim {\hbar \over |\mathbf{q}|}$, but a sharp relation can only really be made in the eikonal regime $J \gg \hbar $ (more on this below) with the details depending on the nature of the interaction.\footnote{For example, in Newtonian gravity at leading order in the Newton's constant $G$ we have $b = {\alpha_G \over |\mathbf{q}|}$ with $\alpha_G = {G m_1 m_2}$ the dimensionless gravitational coupling, where $m_1$ and $m_2$ are the masses of the particles.}
    \item Compton wavelength, $\lambda_{\text{C}}=\frac{\hbar}{m}$, where $m$ is the lightest mass in the QFT, which may be as large as the mass of the lightest scattered particle. This scale is associated with production of massive particles from the vacuum, which is a genuine QFT effect.
  
\end{itemize}
These scales are represented in the Fig. \ref{fig:diagram_scales} below.

We are interested in regimes where QFT effects are suppressed. In particular, we would like to neglect short-range effects due to the virtual exchange of massive particles (such as virtual electron-positron pairs). This is achieved by assuming the separation between the scattered wavepackets to be much larger than any Compton wavelength in the QFT,
\begin{equation}\label{eq:class_b}
    b \gg \lambda_{\text{C}}\, .
\end{equation}

\begin{figure}
    \centering
    \includegraphics[scale=0.4]{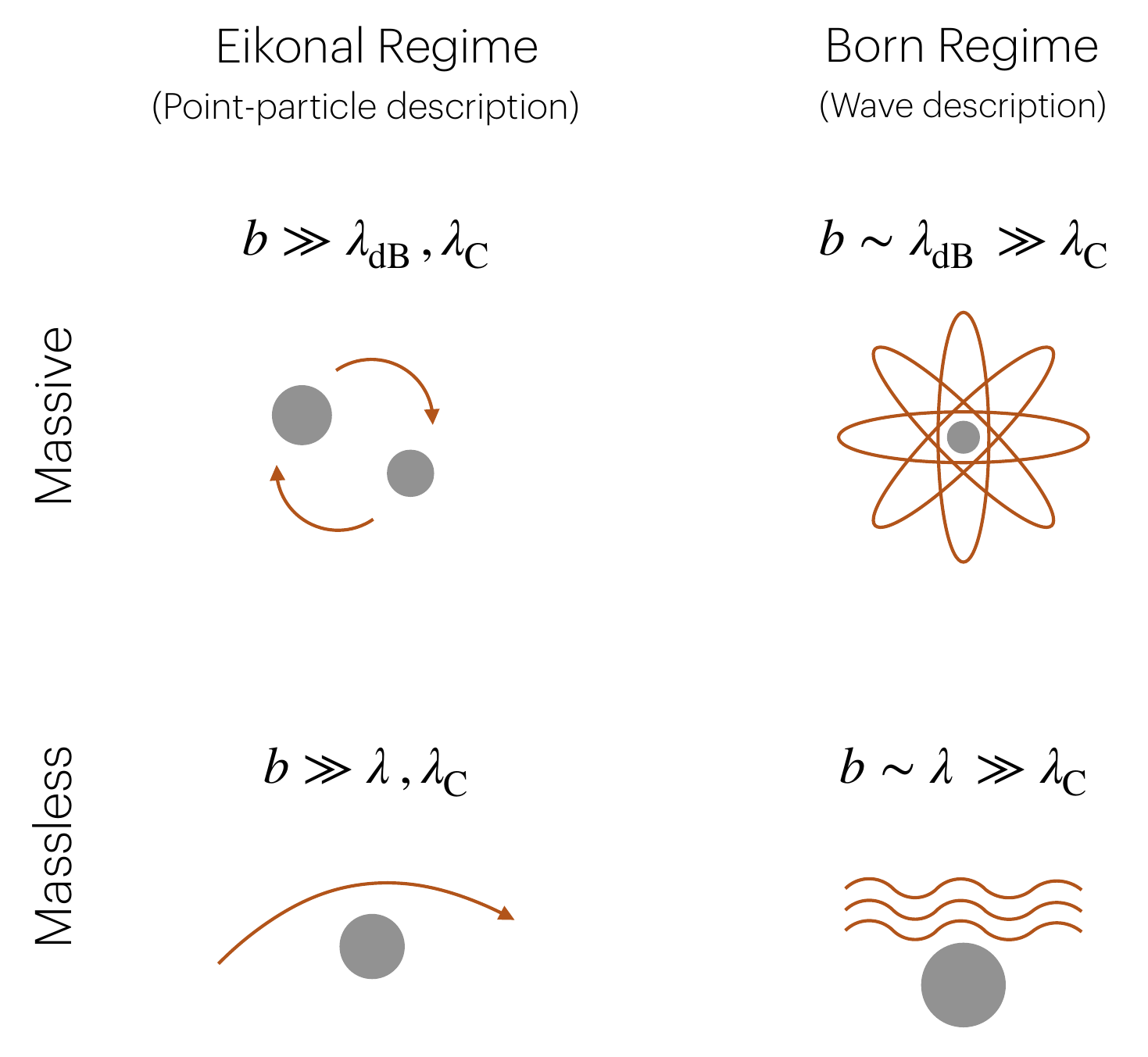}
    \caption{Schematic representation of the different classical regimes considered in this work. The eikonal regime (left hand side), identified with a large impact parameter $b$, and corresponding to classical relativistic mechanics for massive particles or a geometric optics approximation for massless particles. The Born regime (right hand side) characterised by $b\sim \lambda$, which is relevant for quantum mechanics (massive) or classical wave mechanics (massless). }
    \label{fig:regimes-table}
\end{figure}

Given the above condition, we identify four distinct regimes depending on whether the scattering particles are massive or massless and on how the wavelength $\lambda$ compares with $b$ and $\lambda_C$. For convenience of the reader all the regimes are summarized in Fig. \ref{fig:regimes-table}. We will now describe each of these regimes in detail.

\begin{figure}[h]
    \centering
    \includegraphics[scale=0.45]{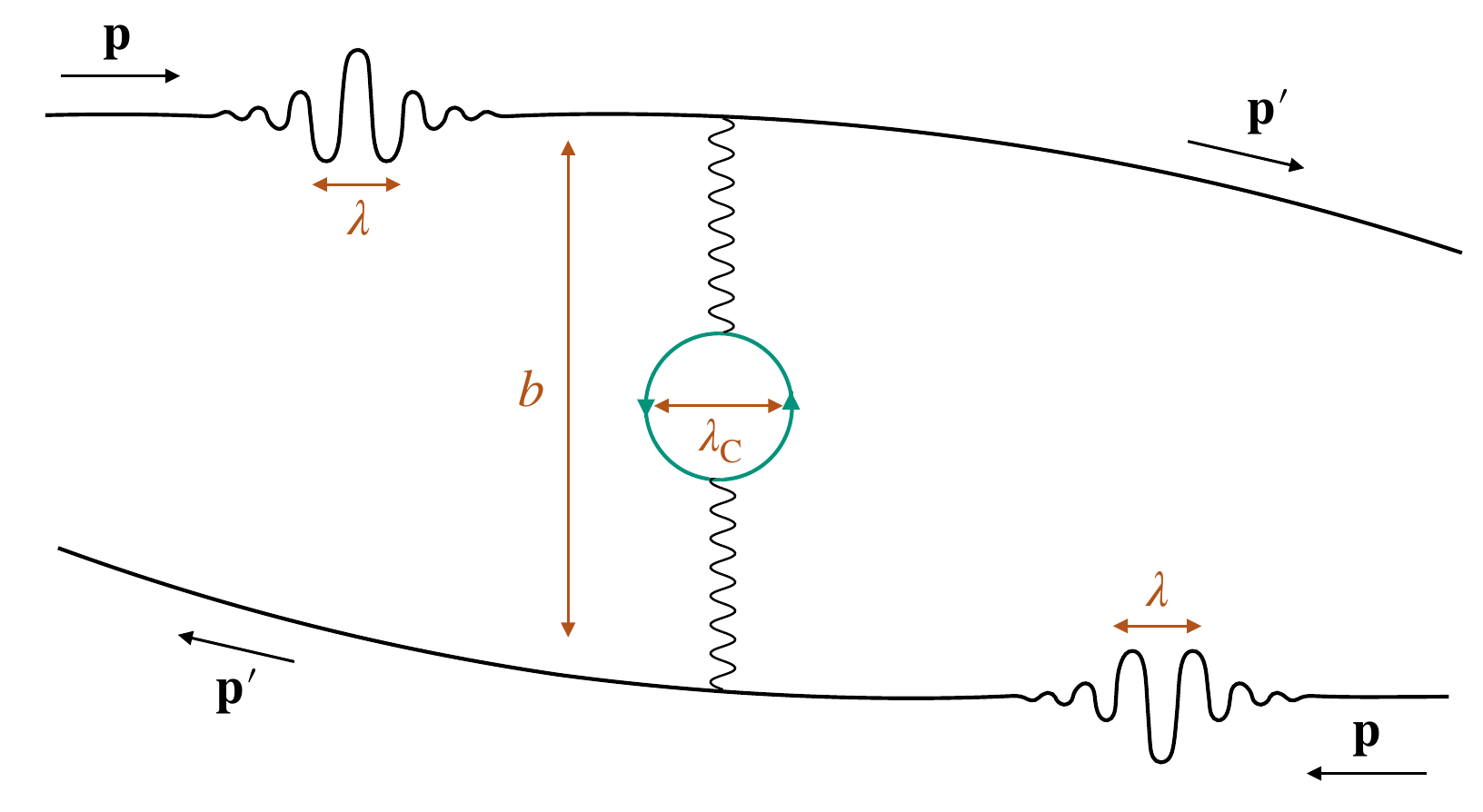}
    \caption{Pictorial representation of the kinematic scales intervening in $2\rightarrow 2$ scattering: the impact parameter $b$ between the two wavepackets, the wavelength of the scattered wavepackets $\lambda$, which reduces to the de Broglie wavelength for massive particles and the ordinary wavelength for massless waves. Lastly, the Compton wavelength $\lambda_\mathrm{C}$, related to the lightest particle in the spectrum (in blue).}
    \label{fig:diagram_scales}
\end{figure}

We first consider both particles to be massive $\lambda = \lambda_\text{dB}$. If the wavepackets are well-separated compared to their wavelengths, $b \gg \lambda_\text{dB}$, a classical point-particle description becomes adequate (top left panel in Fig. \ref{fig:regimes-table}). In terms of the angular momentum $J = b |\mathbf{p}|$ this limit reads $J \gg \hbar$. This regime is known as \emph{eikonal}, and it is where the WKB approximation of quantum mechanics holds. Furthermore, the relationship between $\lambda_\text{dB}$ and $\lambda_C$ will dictate if this classical point-particle description is Newtonian ($\lambda_\text{dB} \gg \lambda_\text{C}$) or relativistic ($\lambda_\text{dB} \sim \lambda_\text{C}$). So we have
\begin{equation}
 \text{Classical (relativistic) mechanics: }  \qquad  b  \gg \lambda_\text{dB}, \, \lambda_\text{C}\, . 
\end{equation}
In this regime the Feynman diagram expansion is observed to `exponentiate' in the impact parameter $b$ space, an observation known as eikonal exponentiation \cite{Amati:1987uf,Amati:1987wq,Amati:1990xe,Amati:1992zb,Kabat:1992tb,Bellazzini:2022wzv,Bern:2021dqo,Herrmann:2021tct,DiVecchia:2021bdo,Brandhuber:2021eyq,Bjerrum-Bohr:2021vuf,DiVecchia:2023frv}. Classical observables such as the scattering angle or the transferred momentum $|\mathbf{q}|$  become determined in terms of $b$ and $|\mathbf{p}|$ via a saddle-point approximation. This is the relevant regime for the inspiral stage of black hole binaries. The majority of the work in using amplitudes methods has been in this regime (see \cite{Buonanno:2022pgc} for a review).

\par 

If we instead let the wavepackets be widespread $\lambda_\text{dB} \sim b$ and requiring QFT effects to be suppressed $b \gg  \lambda_\text{C}$ we must have $\lambda_\text{dB} \gg \lambda_\text{C}$ which implies the non-relativistic limit $|\mathbf{p}| \ll m$. The regime where the Compton wavelength is much smaller than all other length scales is of course the regime where quantum mechanics applies (top right in Fig. \ref{fig:regimes-table}),
\begin{equation}
 \text{Quantum mechanics: }  \qquad  b \sim \lambda_\text{dB} \gg \lambda_\text{C}\,.
 \label{eq:qm}
\end{equation}
In this regime, which we refer to as the \emph{Born regime}, the Feynman diagram expansion has to reduce to the Born series of the Schr\"{o}dinger equation with the appropriate potential. In the gravitational context, this regime is useful to describe axion clouds around black holes, so called gravitational atoms \cite{Arvanitaki:2014wva,Adams:2022pbo,Kabat:1992tb,Adamo:2022ooq,Khalaf:2023ozy}. As is well known, going away from the non-relativistic regime requires including all sorts of QFT effects such as pair production. That is, simply generalizing the Schr{\"o}dinger equation to the Klein-Gordon equation with a \emph{real} potential does not suffice.

Let us now discuss the scales and regimes when one of the particles is massless. In the center of mass frame, the momenta of both particles is the same in absolute value $|\mathbf{p}|$. For a massless particle, corresponding to a plane wave with frequency $\omega$, we have $|\mathbf{p}| = \hbar \omega$ and therefore $\lambda = {\hbar \over |\mathbf{p}|} = {1 \over \omega}$ is the wavelength of the wave associated with the massless particle. 

The discussion proceeds in an analogous way to the massless case. If the wavepacket is narrow compared to the distance $b$ to the massive body such that any uncertainty on the trajectory can be safely neglected \cite{Bellazzini:2022wzv} then the eikonal regime holds, as before. This is the regime of
\begin{equation}
 \text{Geometric optics: }  \qquad  b  \gg \lambda, \, \lambda_\text{C}\, . 
 \label{eq:go}
\end{equation}
In this regime (bottom left in Fig. \ref{fig:regimes-table}), eikonal exponentiation of the amplitude is observed in the same way as for the massive case \cite{Amati:1987uf,Amati:1987wq,Amati:1990xe,Amati:1992zb,Kabat:1992tb,Bellazzini:2022wzv,Bellazzini:2021shn,Bjerrum-Bohr:2017dxw,Bjerrum-Bohr:2016hpa,AccettulliHuber:2020dal,Chen:2022clh}, and classical phenomena such as gravitational lensing and the Shapiro time-delay follow from a saddle-point  localization in a similar way.\footnote{Technically, the regime \eqref{eq:go} only accounts for \emph{classical} geometric optics. If $\lambda_\text{C} \sim b$ while $b \gg \lambda$ then the virtual exchange of massive matter, 
 such as electron-positron pairs, may correct the classical gravitational deflection angle. This leads to a non-universal coupling between different massless particles (say photons vs gravitons), i.e. a quantum violation of the equivalence principle, which in practice is very small in astrophysical contexts \cite{Bellazzini:2021shn}.}

At last, if the wavepacket is extended across space $\lambda \sim b$ with a wavelength much larger than the Compton wavelength, then a classical wave description applies (bottom right in Fig. \ref{fig:regimes-table}),
\begin{equation}
 \text{Classical wave mechanics: }  \qquad  b \sim \lambda \gg \lambda_\text{C} 
 \label{eq:cw}
\end{equation}
This is the massless equivalent of the quantum mechanical regime \eqref{eq:qm}, which we also qualify as a Born regime. The hierarchy between $\lambda \gg \lambda_\text{C}$ in this case implies that $\hbar \omega \ll m$ which requires the \emph{probe} limit for the massless particle \footnote{We employ the term ``probe limit", even though it is conventionally used in the massive case, as the wave does not back-react on the geometry. }. This is a very important fact that can simplify calculations.

In this classical Born regime we expect the Feynman diagram expansion to reduce to the Born series of a classical wave equation.  We can think of scattering in this regime as a spectroscopy experiment, where properties of the compact object are probed by a wave. In gravitational wave physics, this regime is relevant in the ringdown phase of black hole binary coalescence, where the space-time consists of gravitational wave perturbations on top of a single black hole background.  In this work we will use this understanding to compute gravitational amplitudes in this regime (see also \cite{Bautista:2021wfy,Bautista:2022wjf,Ivanov:2022qqt,Ivanov:2024sds} for other work) \footnote{This regime is typically  referred to as ``classical limit" in the  literature, due to the hierarchy  $\hbar \omega \ll m$. Nevertheless, we use the term ``Born regime'' to stress the structure the amplitude takes in terms of a Born series, akin to non-relativistic scattering in quantum mechanics (see Chapter \ref{sec:EOB}).}.

The last possible scenario is the scattering of two massless particles. The eikonal regime of this configuration has been studied in the context of gravity in seminal work on the subject \cite{Amati:1987uf,Amati:1987wq,Amati:1990xe,Amati:1992zb,Kabat:1992tb,Verlinde:1991iu,tHooft:1987vrq}, and corresponds to the propagation of null geodesics on a shockwave background. We expect the related Born regime to describe a classical wave on such background, but we do not discuss this further in this work.

So far we have exclusively discussed kinematic scales, but of course turning on Newton's constant $G$ introduces further dynamical scales. In particular, the Schwarzschild radius $R_s =2 G E_{\mathrm{c.m.}}$ and the Planck's length $\lambda_{\mathrm{Pl}}=\sqrt{G\hbar}$ (see discussion in \cite{Bellazzini:2022wzv}). Since the Schwarzschild radius is the length scale where gravity becomes strongly coupled, the small parameter in a perturbative calculation is $R_s/b$, which in the eikonal regime corresponds to the well-known PM corrections, weighted by the impact parameter between the two particles. Notice that in the massless/massive Born regime \eqref{eq:cw}, we are probing scales of the order of the wavelength $\lambda$ and therefore the small parameter is $\sim R_s/\lambda=2GM\omega$, where we replaced $R_s=2GM$. This statement is equivalent to the fact that the wavelength of the wave probing the background should be larger than the Schwarzschild radius, meaning the black hole can not be fully resolved by the probe and at leading order is accurately described by a point particle. In the BHPT literature this is known as the MST low-frequency expansion \cite{Mano:1996gn,Mano:1996vt}. 

In the case of scattering against a Kerr BH there is the additional length scale associated with the spin $a$ of the BH, for which the sub-extremality bound implies $a < G M$, meaning that corrections in $a$ are always subleading compared to corrections in $G$ at the same order. In this case the low-frequency expansion in BHPT will involve a joint perturbative expansion in $a$ and $G$ given that the dimensionless quantities are $a \omega$ and $G M \omega$. See \cite{Bautista:2023sdf} for a recent discussion.

\begin{figure}[h]
    \centering
    \includegraphics[scale=0.4]{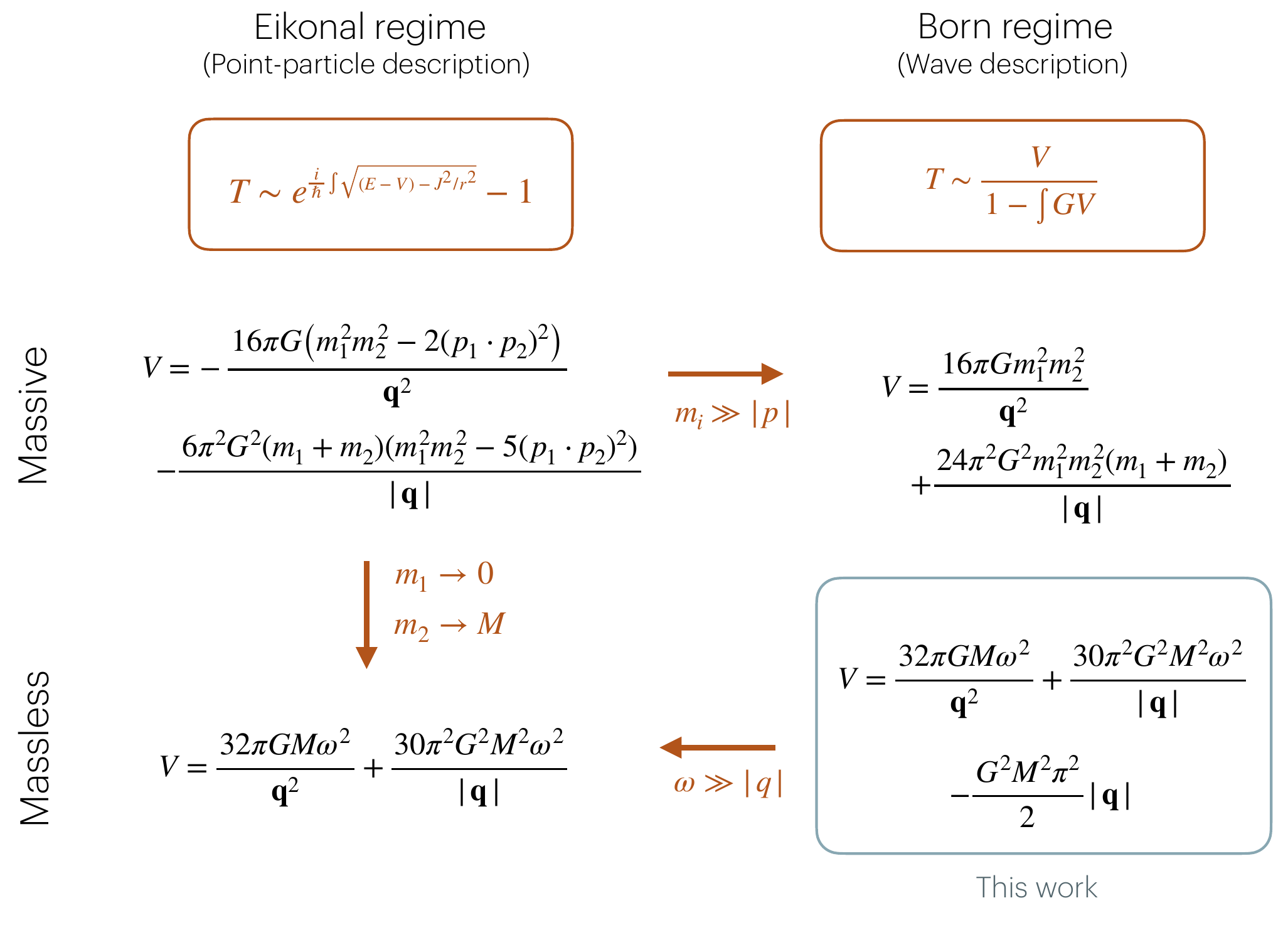}
    \caption{The gravitational potential of two minimally coupled scalars at $O(G^2)$ in various classical regimes. In the eikonal regime the amplitude exponentiates in a classical radial action, while in the Born regime the amplitude is given by a geometric Born series, as qualitatively shown in the orange boxes. The arrows represent how the potentials the in different regimes are accessible between each other.}
    \label{fig:regimes-V}
\end{figure}

Anticipating the discussion in the remaining Chapters, it is useful to introduce the potential $V(\mathbf{p},\mathbf{p}')$, where $\mathbf{p}$ and $\mathbf{p}'$ are the incoming and outgoing momenta in the center of mass frame and $\mathbf{q} \equiv \mathbf{p}' - \mathbf{p}$ is the momentum transfer. The relation between the amplitude and the potential is represented in Figure \ref{fig:regimes-V} depending on the different regimes. In the Born regime the amplitude takes the form a geometric series in terms of the potential which is the Born series. In the eikonal regime, there is exponentiation and the amplitude depends on the potential via the exponent, which is called the radial action. The potential is a interesting object on its own because it also allows to determine bound observables like the perihelion precession of gravitational orbits \cite{Kalin:2019rwq,Kalin:2019inp,Cho:2021arx} or the energy levels of bounds states \cite{Todorov:1970gr, Rizov:1975tr,Crater:1992xp,Jallouli:1996bu}. In Figure \ref{fig:regimes-table} we list the gravitational potential of two minimally coupled scalars at $O(G^2)$ showcasing the differences between the different regimes.

In the next Chapter we will formally define the potential by first introducing the effective-one-body (EOB) framework in QFT. We will discuss how the EOB Schr{\"o}dinger equation reduces in the Born regime to a non-relativistic Schr\"odinger equation (massive) or a classical wave equation (massless).

\section{Effective-one-body framework in Quantum Field Theory}\label{sec:EOB}

From now on we will set natural units $\hbar = c  = 1$.

In this Chapter we will make the discussion of the previous section  more quantitative. Following Todorov \cite{Todorov:1970gr} we will write down a relativistic effective-one-body Lippmann-Schwinger equation that defines a potential in terms of Feynman diagrams. We will then require this Lippmann-Schwinger equation to be consistent with relativistic two-particle unitarity. From there we can define a relativistic EOB Schr{\"o}dinger equation with the potential that is determined from the scattering amplitudes.  The statement then essentially becomes that, in either of Born regimes discussed in Chapter \ref{sec:scales}, the potential becomes analytic in the energy, only having singularities for $t= q^2 = 0$ corresponding to long-distance interactions, just as in quantum mechanics. Therefore the EOB Schr{\"o}dinger equation reduces to the usual Schr{\"o}dinger equation of quantum mechanics in case \eqref{eq:qm} or a classical wave equation in case \eqref{eq:cw}. We start by establishing the EOB kinematics.

\subsection{Effective-one-body kinematics}\label{sec:Todorov}

In the center of mass (CM) frame we have the ingoing momenta
\begin{align}\label{eq:kinp1p2}
p_1^\mu &= (E_1, \mathbf{p})\, , & E_1 = \sqrt{|\mathbf{p}|^2 + m_1^2}\, , \notag\\
p_2^\mu &= (E_2, -\mathbf{p})\, , & E_2 = \sqrt{|\mathbf{p}|^2 + m_2^2}\, , 
\end{align}
with the outgoing momenta given by
\begin{align}\label{eq:kinp3p4}
    p_3^\mu &= (E_1, \mathbf{p}') = (E_1, \mathbf{p}+ \mathbf{q})\, , \notag \\
p_4^\mu &= (E_2, -\mathbf{p}') =  (E_2, -\mathbf{p} - \mathbf{q})\, .
\end{align}
Here, $\mathbf{q} = \mathbf{p}' - \mathbf{p}$ is the momentum exchange. 

The Mandelstam invariants read
\begin{equation}
    s = (p_1 + p_2)^2, \qquad t = (p_3 - p_1)^2\, .
\end{equation}
These relate to $\mathbf{p}$ and $\mathbf{q}$ in the CM frame via
\begin{equation}
\label{eq:st}
    |\mathbf{p}|^2 = {\big[s - (m_1+m_2)^2\big]\big[s - (m_1-m_2)^2\big] \over 4 s}, \qquad  t =  -|\mathbf{p}' - \mathbf{p}|^2 = -|\mathbf{q}|^2 .
\end{equation}
The vectors $\mathbf{p}$ and $\mathbf{q}$ are further constrained by the condition $\mathbf{p} \cdot \mathbf{q} = - |\mathbf{q}|^2 / 2$, that stems from energy conservation $|\mathbf{p}| = |\mathbf{p}'|$.

\par
We will interpret $\mathbf{p}$ and $\mathbf{p}'$ as the incoming and outgoing momentum of \emph{one} particle that is scattering off some potential (which we will define in an instant). In terms of the kinetic energy $E$ this effective one body (EOB) will satisfy a non-trivial energy-momentum relation given by 
\begin{equation}
\label{eq:emeob}
    |\mathbf{p}|^2 = {E (E + 2 m_1)(E + 2 m_2)(E + 2 m_1 + 2 m_2)\over 4 (E + m_1 + m_2)^2}
\end{equation}
where we replaced $E \equiv \sqrt{s} -  (m_1 + m_2)$ in   \eqref{eq:st}. 
Taking the non-relativistic limit $E \to 0$ we find 
\begin{align}
     |\mathbf{p}|^2 =  2 \mu E\, , \quad &\text{ in the quantum mechanical Born regime \eqref{eq:qm}},\notag\\&\text{ with } \quad \mu = {m_1 m_2 \over m_1 + m_2},
     \label{eq:nonrel}
\end{align}
which is the well-known non-relativistic energy-momentum relation for a body with reduced mass $\mu$. 

Another simple consistency check is to consider one of the particles to be very heavy, say $m_2 \gg m_1, E$. Then,
\begin{equation}
    |\mathbf{p}|^2 = E^2 + 2 m_1 E = E_1^2 - m_1^2,
\end{equation}
where $E_1 = E + m_1$ in this limit. Thus, when $m_2$ is very heavy, the effective one-body is in fact just the one body $m_1$. 

In the Born regime for the scattering of a classical wave \eqref{eq:cw}, we further have the probe limit $m_1/m_2 \to 0$, in which case
\begin{equation}
    |\mathbf{p}|^2 = E^2 = \omega^2, \text{ in the classical Born regime \eqref{eq:cw}}
    \label{eq:cw2}
\end{equation}

\subsection{Born series as a solution to relativistic unitarity}

Let $T(s,t)$ be the scattering amplitude for the scattering of two scalars of mass $m_1$ and $m_2$. We will use the notation $T(\mathbf{p}, \mathbf{p}')$ where it is understood that the dependence on $s$ and $t$ is given in terms of the initial and final momentum, $\mathbf{p}$ and $\mathbf{p}'$, in \eqref{eq:st}.

We want to interpret $T(\mathbf{p}, \mathbf{p}')$ as the amplitude of scattering off of some potential $V(\mathbf{p}, \mathbf{p}')$ where $\mathbf{p}$ and $\mathbf{p}'$ are, respectively, the initial and final momentum of the EOB, which respect the energy-momentum relation in \eqref{eq:emeob}.

The relation between the amplitude $T(\mathbf{p}, \mathbf{p}')$ and the potential $V(\mathbf{p}, \mathbf{p}')$  is given by the Lippmann-Schwinger equation, which serves as a definition of $V(\mathbf{p}, \mathbf{p}')$,
\begin{equation}
    \adjustbox{valign=c,scale={1}{1}}{
\includegraphics[scale=0.25]{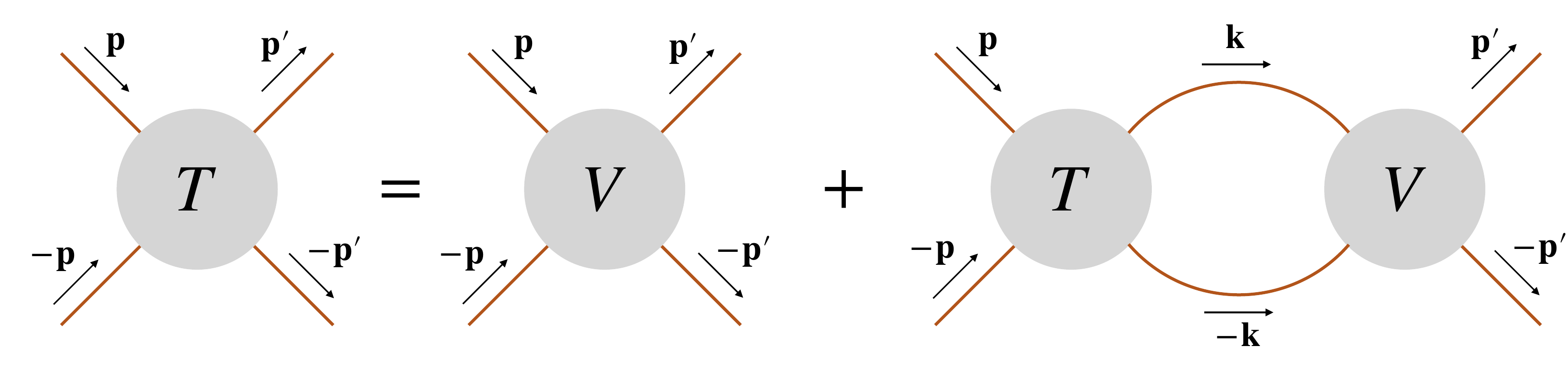}
    } 
    \label{eq:LS-diagram}
\end{equation}
or, explicitly,
\begin{equation}\label{eq:LS}
    T(\mathbf{p},\mathbf{p}^\prime)=V(\mathbf{p},\mathbf{p}^\prime)+\int \d^3\mathbf{k}\,T(\mathbf{p},\mathbf{k})G(\mathbf{p},\mathbf{k})V(\mathbf{k},\mathbf{p}^\prime)\, ,
\end{equation}
where $G(\mathbf{p},\mathbf{k})$ is a Green's function which will be fixed shortly by requiring consistency of \eqref{eq:LS} with (relativistic) unitarity. Note that in the last term the integrated momentum $\mathbf{k}$ can be off-shell.\footnote{We therefore make the assumption that the scattering amplitude $T(\mathbf{p}, \mathbf{p}')$ admits a good analytic continuation away from the mass-shell. This is of course verified in perturbation theory (see \cite{Cutkosky, Correia:2022dcu} for a discussion in the nonperturbative case).}

The fact that we chose the integration in \eqref{eq:LS} to be 3-dimensional is to make direct contact with the standard Lippmann-Schwinger equation of one-body quantum mechanics.\footnote{The Lippmann-Schwinger equation can also be written covariantly (see \cite{Todorov:1970gr}).}
As in quantum mechanics, we would like the potential $V(\mathbf{p},\mathbf{p}') = V(s,t)$ to be a real function. This would imply that the imaginary part of $T(\mathbf{p},\mathbf{p}')$ would have to come from the iteration piece in \eqref{eq:LS} involving the Green's function $G(\mathbf{p},\mathbf{k})$. However, in quantum field theory, scattering amplitudes have several branch cuts, not just on the elastic threshold of $m_1$ and $m_2$, but also on inelastic thresholds for particle production (which are typically present at higher energies). 

We therefore let the potential $V(\mathbf{p},\mathbf{p}')$ be real across the elastic two-particle threshold (2PC) of $m_1$ and $m_2$ but imaginary across possible inelastic thresholds,
\begin{equation}
\mathrm{Im}\, V(\mathbf{p},\mathbf{p'})|_\text{2PC} = 0.
\label{eq:ImV}
\end{equation}
We will now require consistency of the Lippmann-Schwinger equation \eqref{eq:LS} with elastic unitarity, i.e. unitarity across the 2PC, in order to fix the form of $G(\mathbf{p},\mathbf{k})$. Elastic unitarity reads \cite{Correia:2020xtr}
\begin{align}
\mathrm{Im} \, T(\mathbf{p},\mathbf{p}')|_\text{2PC} &= {1 \over 8 \pi^2} \int {\d^4 k} \; \delta^+\!\big(k^2 - m_1^2\big) \delta^+\!\big((k-p_1 - p_2)^2 - m_2^2 \big) T(p,k) \, T^*(k,p') \notag \\
&= {1 \over 16 \pi^2 \sqrt{s}} \int {\d^3 \mathbf{k} } \, \delta(|\mathbf{k}|^2 - |\mathbf{p}|^2) \, T(\mathbf{p},\mathbf{k}) \, T^*(\mathbf{k},\mathbf{p}')\, ,
\label{eq:elauni}
\end{align}
where $k$ is the 4-momentum and $\delta^+$ puts the corresponding particle on-shell and imposes the positive-energy condition, $\delta^+\!\big(k^2 - m_1^2\big) \equiv \Theta(k^0) \delta\!\big(k^2 - m_1^2\big)$. In the second line, we integrated over $k^0$ in the center of mass frame to have a three-dimensional integration as in the Lippmann-Schwinger equation \eqref{eq:LS}.

To arrive at the constraint on $G$ we will work with a short-hand notation. Schematically, the Born series, which is a formal solution to the Lippmann-Schwinger equation \eqref{eq:LS} reads
\begin{equation}\label{eq:T-Born}
    T = V + V G V + V G V G V + \dots = V {1 \over 1 - G V}\, ,
\end{equation}
where we suppressed explicit kinematic dependence and integration symbols.

Imposing the reality condition \eqref{eq:ImV} across the 2PC ($V^* = V$  in short-hand notation) gives
\begin{equation}\label{eq:T-star}
    T^* = V {1 \over 1 - G^* V}\, .
\end{equation}
In the same notation, elastic unitarity \eqref{eq:elauni} takes the form
\begin{equation}
    T - T^* = T T^* \, .
\end{equation}
Using \eqref{eq:T-Born} and \eqref{eq:T-star}, the right-hand side can be written as
\begin{equation}
    T T^* = V {1 \over 1 - G V} V {1 \over 1 - G^* V} \, ,
\end{equation}
whereas the left-hand side is
\begin{equation}
    T - T^* = V \left[ {1 \over 1 - G V} - {1 \over 1 - G^* V} \right] = V {1 \over 1 - G V} (G - G^*) V {1 \over 1 - G^* V} \, .
\end{equation}
Unitarity then fixes $G - G^* = 1$, which explicitly reads
\begin{equation}
    \mathrm{Im} \, G(\mathbf{p}, \mathbf{k}) ={\delta(|\mathbf{k}|^2 - |\mathbf{p}|^2) \over 16 \pi^2 \sqrt{s}}  \, .
    \label{eq:ImG}
\end{equation}
This constraint is satisfied by the Green's function
\begin{equation}
    G(\mathbf{p},\mathbf{k}) = {1 \over (2 \pi)^3 } {1 \over 2 \sqrt{s}}  {1 \over |\mathbf{k}|^2 - |\mathbf{p}|^2 - i \epsilon}\, ,
    \label{eq:G}
\end{equation}
where the sign of $i \epsilon$ is chosen such that the scattering amplitude on the physical region is defined above the cut, $T_\text{s-channel}(s,t) = T(s + i \epsilon,t)$.\footnote{In particular, we have that $s \to s + i \epsilon$ implies $|\mathbf{p}|^2 \to |\mathbf{p}|^2 + i \epsilon \left({s^2 - (m_1 - m_2)^2 \over 4 s^2} \right)$, where $\left({s^2 - (m_1 - m_2)^2 \over 4 s^2} \right) > 0$ on the physical region $s \geq (m_1 + m_2)^2$. Note that the Feynman $i \epsilon$ prescription sets the Green's function to be \emph{retarded}.} 

We could have also added analytic pieces to \eqref{eq:G}, which drop out when taking the imaginary part. These additional pieces however would lead to worse UV behavior $|\mathbf{k}| \to \infty$ of the integrand in \eqref{eq:LS}, so we chose not to include them.

In the non-relativistic limit \eqref{eq:nonrel} we indeed recover the `propagator' of the Born series in quantum mechanics \cite{Sakurai:2011zz},
\begin{align}
    G(\mathbf{p},\mathbf{k}) \to {1 \over (2 \pi)^3} {1 \over 4 m_1 m_2} \, {1 \over {|\mathbf{k}|^2 \over 2 \mu} -  E  - i \epsilon}\,,  \,\!\! \text{ in the quantum mechanical Born regime \eqref{eq:qm}}\,,
    \label{eq:Gqm}
\end{align}
where $E$ is the kinetic energy, as defined in \eqref{eq:emeob}.

In the regime relevant for classical wave scattering \eqref{eq:cw}, where  $m_1 = 0$ and $m_2  = M \to \infty$, we find instead
\begin{equation}
    G(\mathbf{p},\mathbf{k}) \to {1 \over (2 \pi)^3} {1 \over 2 M}\, {1 \over {|\mathbf{k}|^2} -  \omega^2  - i \epsilon}\, , \, \text{ in the classical wave Born regime \eqref{eq:cw}}
    \label{eq:Gcw}
\end{equation}
where $\omega = E = |\mathbf{p}|$ is the frequency of the massless particle.

\subsection{Effective-one-body Schr{\"o}dinger equation}

 Note that the inverse of the Green's function $G(\mathbf{p},\mathbf{k})$ in \eqref{eq:G} depends quadratically on $\mathbf{k}$,  exactly as  in quantum mechanics \eqref{eq:Gqm}. This implies that \eqref{eq:G}  is the Green's function of a second-order differential equation. Concretely,
 \begin{equation}
    (\nabla^2+|\mathbf{p}|^2) \,G(\mathbf{x},\mathbf{x}^\prime) = \delta^3(\mathbf{x}-\mathbf{x}^\prime)\,,
\end{equation}
where 
\begin{equation}\label{eq:GX}
    G(\mathbf{x},\mathbf{x}^\prime) \equiv - {1 \over 4 \pi} {e^{i |\mathbf{p}||\mathbf{x} - \mathbf{x}'|} \over |\mathbf{x} - \mathbf{x}'|}  = - 2 \sqrt{s} \int \d^3 \mathbf{k} \, e^{i \mathbf{k} \cdot \mathbf{x}} \, G(\mathbf{p},\mathbf{k}) \,,
\end{equation}
with $G(\mathbf{p},\mathbf{k})$ given by \eqref{eq:G}.

Now, we let $\Psi(\mathbf{x})$ be the effective-one-body wavefunction. It obeys the effective-one-body Schr{\"o}dinger equation:
\begin{equation}
     (\nabla^2+|\mathbf{p}|^2) \Psi(\mathbf{x})= V(\mathbf{x}) \Psi(\mathbf{x})\,,
     \label{eq:EOBSchrodinger}
\end{equation}
where $|\mathbf{p}|$ is given in terms of the energy $E$, according to the non-linear relation \eqref{eq:emeob}. 

In the above, $V(\mathbf{x})$ is the potential in position space, which we define via the Fourier transform
\begin{equation}
    V(\mathbf{p},\mathbf{p}')= -2\sqrt{s}\int \!\d^3 \mathbf{x} \, e^{-i  \mathbf{p}'\cdot \mathbf{x}} \, V(\mathbf{x}) \, e^{i  \mathbf{p}\cdot \mathbf{x}}\,,
    \label{eq:VEOB}
\end{equation}
where $\mathbf{p}$ is the incoming EOB momentum, and $\mathbf{p}'$ is the outgoing EOB momentum, with $|\mathbf{p}'| = |\mathbf{p}|$ given by \eqref{eq:emeob} in terms of the energy $E$. Note that $V(\mathbf{x})$ is in general an operator and may contain derivatives, that can both act on the right and on the left (see Chapter \ref{sec:Kerr}).

With these definitions, the EOB Schr{\"o}dinger equation \eqref{eq:EOBSchrodinger} reads in momentum space,
\begin{equation}
\label{eq:eobms}
   \Psi(\mathbf{p}') = G(\mathbf{p},\mathbf{p}') \int \! \d^3 \mathbf{k} \; V(\mathbf{k},\mathbf{p}') \Psi(\mathbf{k})\,,
\end{equation}
where $\Psi(\mathbf{k}) \equiv \int \d^3 \mathbf{x} \, e^{-i  \mathbf{k}\cdot \mathbf{x}} \, \Psi(\mathbf{x}) $.

Proceeding as in textbook quantum mechanics \cite{Sakurai:2011zz}, we look at the 
 inhomogenous solution to \eqref{eq:EOBSchrodinger} in order to determine the scattering of an `incoming' plane wave $\Psi_{\mathbf{p}}^{(0)}(\mathbf{x}) = e^{i \mathbf{p} \cdot \mathbf{x}}$ against the potential $V(\mathbf{x})$. The solution can be expressed explicitly in terms of $\Psi_{\mathbf{p}}^{(0)}(\mathbf{x})$ via the integral relation
\begin{align}
    \Psi_{\mathbf{p}}(\mathbf{x})&=\Psi_{\mathbf{p}}^{(0)}(\mathbf{x})+\int \d^3 \mathbf{x}^\prime \, G(\mathbf{x},\mathbf{x}^\prime) \,V( \mathbf{x}^\prime) \, \Psi_{\mathbf{p}}(\mathbf{x}^\prime)\,.
\end{align}

Finally, the scattering amplitude $T(\mathbf{p}, \mathbf{k})$ is defined as the coefficient of the `outgoing' wave at infinity $\Psi_{\mathbf{p}}(\mathbf{x} \to \infty)$,
\begin{equation}\label{eq:TEOB}
    T(\mathbf{p}, \mathbf{p}') = - 2 \sqrt{s} \int \d^3 \mathbf{x}^\prime \, e^{-i \mathbf{p}' \cdot \mathbf{x}^\prime } \, V(\mathbf{x}^\prime) \,\Psi_{\mathbf{p}}(\mathbf{x}')\,,
\end{equation}
where the overall normalization was fixed such that $T(\mathbf{p}, \mathbf{p}')$ obeys the Lippmann-Schwinger equation \eqref{eq:LS}. We refer the reader to Appendix \ref{app:LS} for details.

Let us briefly recap what we have achieved so far. 
We have reorganized the Feynman diagram expansion in terms of a Born series of the effective-one-body relativistic Schr{\"o}dinger equation given in \eqref{eq:EOBSchrodinger}. The potential $V(\mathbf{x})$ that enters the EOB Schr{\"o}dinger equation is determined by the relativistic scattering amplitude $T(\mathbf{p},\mathbf{k})$ via the Lippmann-Schwinger equation \eqref{eq:LS} and the Fourier transform \eqref{eq:VEOB}. In practice, $V(\mathbf{x})$ can be determined perturbatively by matching the Born series with the Feynman diagram expansion. 

For example, at one-loop in quantum electrodynamics, we have the vacuum polarization diagram, the triangle vertex correction that contributes to the anomalous magnetic moment of the electron, among other diagrams such as the box and cross-box, which are known to contribute to the Lamb shift in the Hydrogen energy levels. These can be determined by fixing the potential $V(\mathbf{x})$ and computing the corrections as in textbook quantum mechanics \cite{Rizov:1975tr,Crater:1992xp,Jallouli:1996bu,Berestetskii:1982qgu}. In practice, the box and cross-box diagrams are the ones that match with the ``iteration"  $\sim V G V$ in the Born series (see worked out example in Appendix \ref{app:iteration}).\footnote{This is an important difference with respect to other proposals such as the Bethe-Salpeter equation \cite{PhysRev.84.1232}, where only the box diagram contributes to the iteration and the cross-box goes into the corresponding ``potential". This makes the Bethe-Salpeter not reduce properly in the probe limit (see \cite{Todorov:1970gr} for further discussion).}

In the more recent context of the gravitational amplitudes program, a similar idea of matching to an effective-field-theory of the two fields that represent the two scattering bodies $m_1$ and $m_2$ was introduced by Cheung, Rothstein and Solon \cite{Cheung:2018wkq}. In that work, the precise matching to the QFT Feynman diagram expansion is done via the so-called $Y$-pole, which is nothing but the EOB Green's function $G(\mathbf{p},\mathbf{p}')$ in \eqref{eq:G}. 

It was noted in \cite{Kalin:2019rwq,Bjerrum-Bohr:2019kec,Cristofoli:2019neg,Cristofoli:2020uzm,Neill:2013wsa} that the amplitude defined modulo iterations of the Y-pole, i.e. the EOB potential $V(\mathbf{p},\mathbf{p}')$ and its Fourier transform  $V(\mathbf{x})$, obey interesting properties. In particular, the EOB potential $V(\mathbf{x})$ satisfies the so-called \emph{impetus} relation $|\mathbf{p}|^2 = |\mathbf{p}(\mathbf{x})|^2 + V(\mathbf{x})$, relating the momentum at a finite ``distance" $\mathbf{p}(\mathbf{x})$ to the potential $V(\mathbf{x})$ and the momentum at infinity $ |\mathbf{p}| \equiv |\mathbf{p}(\mathbf{x} \to \infty)|$ which depends non-linearly on the energy $E$ via \eqref{eq:emeob}. The impetus relation is of course the relativistic EOB generalization of the Newtonian energy conservation between potential and kinetic energies. Relativistic classical observables, including bound orbit observables such as the precession of the perihelion, can then be determined in a similar way to Newtonian mechanics.\footnote{Up to some unresolved limitations due to the causal nature of relativistic interactions. In particular at $O(G^4)$, the EOB potential $V(\mathbf{x})$ gets non-local in time contributions, so-called tail effects, which prevent analytic continuation from $E > 0$, where scattering occurs, to the regime of bound orbits where $E < 0$ (see \cite{Dlapa:2024cje} for a recent account).} 

The impetus relation follows from the EOB Schr{\"o}dinger equation \eqref{eq:EOBSchrodinger} via the WKB approximation, which becomes exact in the eikonal regime $J \gg \hbar$, just as in textbook quantum mechanics \cite{Sakurai:2011zz}. In this regime, the amplitude in partial wave space $J$ exponentiates in terms of the radial action $I_r = \int |\mathbf{p}(r)| dr = \int \sqrt{|\mathbf{p}|^2 -  V(r) - {J^2 \over r^2} } \, dr$, where $r \equiv |\mathbf{x}|$. This is known as the \emph{amplitude-action} relation (see \cite{Bern:2021dqo, Kol:2021jjc} for details).

In this work, however, we are interested in the Born regime. When both particles are massive, $m_1 \neq 0$ and $m_2 \neq 0$, the Born regime is the non-relativistic limit where quantum mechanics holds \eqref{eq:qm}. In this case, the EOB energy-momentum relation \eqref{eq:emeob} becomes the Newtonian relation \eqref{eq:nonrel} and naturally the EOB Schr{\"o}dinger equation \eqref{eq:EOBSchrodinger} reduces to the usual Schr{\"o}dinger equation.

The case of interest here is when one of the particles is massless, where the Born regime is the regime of classical wave scattering \eqref{eq:cw}. In this scenario, the EOB energy-momentum relation \eqref{eq:emeob} reduces to $|\mathbf{p}|^2 = \omega^2$, where $\omega$ is the frequency of the massless particle, and the EOB Schr{\"o}dinger equation \eqref{eq:EOBSchrodinger} takes the form a wave equation,
\begin{equation}
     (\nabla^2+\omega^2) \Psi(\mathbf{x})= V(\mathbf{x}) \Psi(\mathbf{x})\, .
     \label{eq:cweq}
\end{equation}
To argue that this is a \emph{classical} wave equation we need to say more about the potential $V(\mathbf{x})$ in the Born regime. Among other properties, we expect $V(\mathbf{x})$ to be a classical potential if it is a real function and obeys a power law decay with distance. We will now see to what extent this is true in the Born regime.

\subsection{Reduction to a classical wave equation in the Born regime}

It is straightforward to trade statements between $V(\mathbf{x})$ and the Fourier transform $V(\mathbf{p},\mathbf{p}')$, so we will focus instead on the latter since it connects directly to the amplitude $T(\mathbf{p},\mathbf{p}') = T(s,t)$ via the Lippmann-Schwinger equation \eqref{eq:LS}.

Concretely, the nearest singularities in $t = - |\mathbf{p}' - \mathbf{p}|^2$ determine the large distance behavior of the potential $V(\mathbf{x} \to \infty)$. Since the analytic structure of $V(s,t)$ in $t$ is inherited from that of $T(s,t)$, the singularities of $V(s,t)$ in $t$ are determined by the spectrum of particles that can be exchanged in the $t$-channel. The exchange of massless particles, such as the graviton, leads to singularities at $t = 0$. On the side of $V(\mathbf{x})$, the exchange of massless particles implies power-law decay with the distance $|\mathbf{x}|$, as expected from long-range interactions.

\begin{figure}[h]
    \centering
    \includegraphics[scale=0.45]{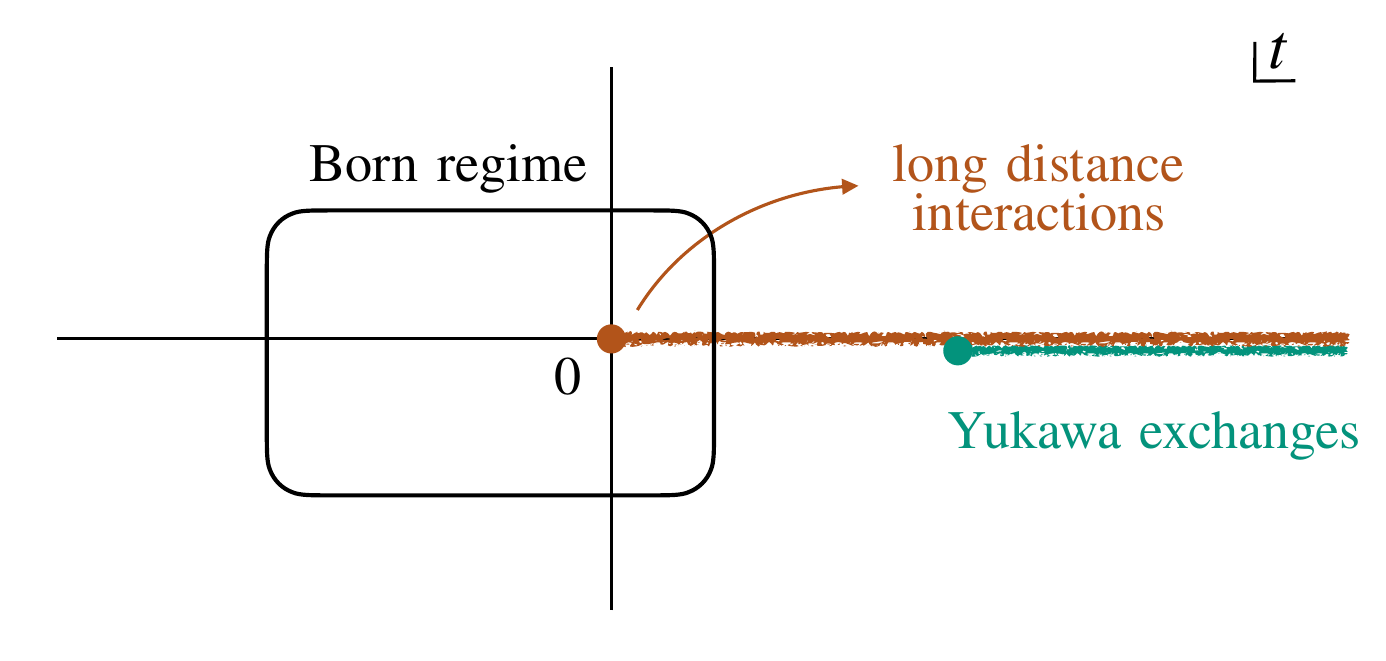}
    \caption{Analytic structure in $t$ in the Born regime. In orange, branch cuts associated with the virtual exchange of massless particles, which give rise to long-range interactions. In blue, branch cuts associated with the virtual exchange of massive particles. These give rise to exponentially decaying contributions to the potential which are unobservable in the Born regime.}
    \label{fig:tplane}
\end{figure}

In a quantum field theory, we can also have all sorts of massive particles being exchanged in the $t$-channel. For example, a singularity at $t = m^2$ gives rise to a Yukawa-like exponential decay $V(\mathbf{x}) \to e^{- m |\mathbf{x}|}$ (see e.g. \cite{Correia:2020xtr, Bellazzini:2021shn}). This is precisely how the vacuum polarization diagram in quantum electrodynamics contributes to the electric potential (the so-called Uehling term \cite{Berestetskii:1982qgu}). However, in the Born regime \eqref{eq:cw} the impact parameter, or the typical distance between the two bodies $\sim |\mathbf{x}|$, is much bigger than the Compton wavelength of any particle, $|\mathbf{x}| \gg \lambda_\text{C}$, which implies $m |\mathbf{x}| \gg 1$. Therefore, virtual exchanges of massive particles in the $t$-channel are suppressed in the Born regime, and no Yukawa-like behavior is observable in the potential $V(\mathbf{x})$. This statement is represented on the analyticity domain in $t$ in Figure \ref{fig:tplane}.

\begin{figure}[h]
    \centering
    \includegraphics[scale=0.45]{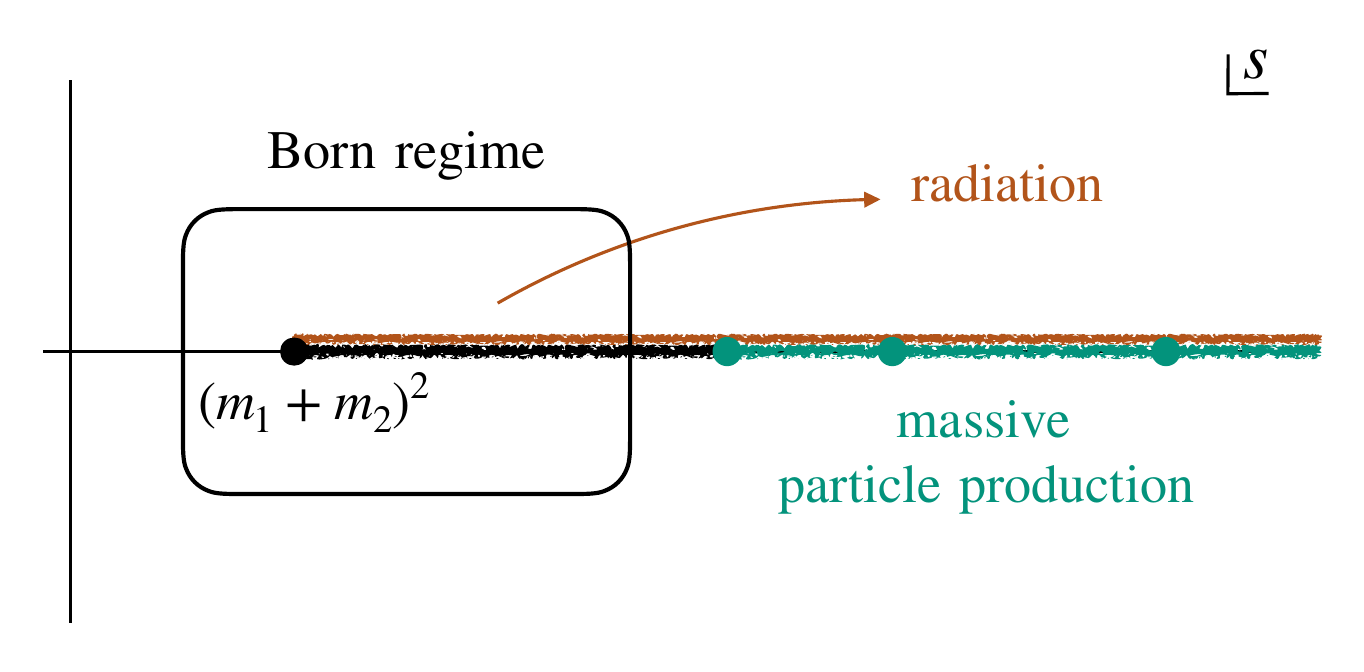}
    \caption{Analytic structure in $s$ in the Born regime. In orange, branch cuts associated with the emission of massless particles, such as classical radiation.  In blue, branch cuts associated with the production of massive particles. In black, the two-particle cut of $m_1$ and $m_2$ contributing to the amplitude but not the potential. In the Born regime, the potential only has branch cuts due to radiation emission, or excitation of other massless degrees of freedom.}
    \label{fig:splane}
\end{figure}

Let us now take a look at the analytic structure in $s$. The potential $V(s,t)$ shares all the singularities of $T(s,t)$ except the elastic two-particle cut across $m_1$ and $m_2$, as defined in \eqref{eq:ImV}. In the Born regime  we have $\lambda \gg \lambda_\text{C}$ which implies $|\mathbf{p}| \ll m$, where $m$ is any mass in the quantum field theory. The relation between $|\mathbf{p}|$ and $s$ in \eqref{eq:st}, implies that any threshold at a finite energy above $s > (m_1 + m_2)^2$ is not accessible in the Born regime. The analyticity domain in $s$ in the Born regime is represented in Figure \ref{fig:splane}.

However, the production of massless quanta in the $s$-channel may be present, corresponding to additional branch cuts on top of the already present two-particle cut $s = (m_1 + m_2)^2$. These additional singularities have to be also in the potential $V(s,t) = V(\mathbf{p},\mathbf{p}')$, which make it develop an imaginary part. This is to be expected, since the emission of radiation is a dissipative process and it is therefore natural that the potential $V(\mathbf{x})$ acquires an imaginary part in this case.\footnote{The branch points may even be below the two-particle cut of course. For example, the diagram giving rise to positronium decay is a box diagram involving the exchange of two-photons in the $s$-channel. This diagram has a branch cut starting at $s = 0$ leading to an imaginary part in $V(\mathbf{x})$, which can be used to compute the decay rate of positronium \cite{Berestetskii:1982qgu}.}

In the classical wave Born regime \eqref{eq:cw}, where one of the scattered particles is massless, we can say more. In this case the condition $\lambda \gg \lambda_\text{C}$ also implies the probe limit $\omega \ll  M$ where $\omega$ is the frequency of the massless particle and $M = m_2$ is the mass of the other scattering particle. The probe limit means that the massless particle $\omega$ does not generate a gravitational field of its own, and therefore its dynamics do not backreact on the geometry. Thus, in the classical wave Born regime \eqref{eq:cw} no emission of gravitational radiation occurs.\footnote{Emission of gravitational radiation is suppressed even when $m_1 \neq 0$, as long as the probe limit $m_1 \ll m_2$ holds. For example, in the eikonal regime $b \gg \lambda$, where a classical point-particle description holds, the lighter particle effectively follows a timelike or null (if $m_1 = 0$) geodesic in a `background' created by the heavier particle $m_2$, without any emission of radiation. Radiation effects (including self-force corrections) are subleading in $m_1/m_2$ (see e.g. \cite{Kosmopoulos:2023bwc,Cheung:2023lnj}).}

These considerations allow us to conclude that in a theory where all interactions are mediated by gravity, in the classical wave Born regime \eqref{eq:cw} no inelastic thresholds are present and the (gravitational) potential has to be analytic in $s$ (or $\omega$). Its only branch points are in $t$ and at the origin, $t = -|\mathbf{q}|^2 = 0$, corresponding to long-distance classical gravitational interactions.
Other interactions associated with the excitation of additional massless degrees of freedom, for example due to horizon absorption or tidal heating of the BH, will add further non-analyticities in $s$ to $T(s,t)$ (see e.g. \cite{Jones:2023ugm,Ivanov:2024sds}) and, as a consequence, an imaginary part to $V(\mathbf{x})$.

Let us now confirm this analysis at $O(G^2)$ in the case of gravitational scattering of two minimally coupled scalars in the Born regime \eqref{eq:cw}.

\section{The gravitational Born regime at 2PM}\label{sec:2PM}
In this Chapter, we carry out an explicit example of the Born regime at one loop: a scalar wave propagating on a Schwarzschild background. The purpose of this example is two-fold: first it will serve as proof of concept of the claims of the previous Chapters, in particular making manifest the emergence of the Born series from scattering amplitudes. At one-loop the series will include iterations of the leading $G$ potential as well as new contributions at $G^2$ (see section \ref{sec:Todorov}). Furthermore, we stress that the computational approach taken here follows very closely usual PM calculations, showing that the Born regime is fully under perturbative control, and can be reconstructed order by order in $G$. This is analog to the Post-Minkowskian expansion relevant for the eikonal regime, which can be readily recovered by taking the high frequency limit of the Born result as discussed in Chapter \ref{sec:scales}.

In the upcoming section, we perform the one-loop calculation in the Born regime $M\gg\omega\sim |\mathbf{q}|$, following \cite{Bjerrum-Bohr:2017dxw,Bjerrum-Bohr:2016hpa,Bern:2019crd,Cheung:2018wkq} and in section \ref{sec:RW} we show how the result precisely matches  its GR counterpart coming from the Regge-Wheeler equation for a scalar wave on a black hole background.

\subsection{One loop scalars in the Born regime}
\begin{figure}
    \centering
    \includegraphics[scale=0.55]{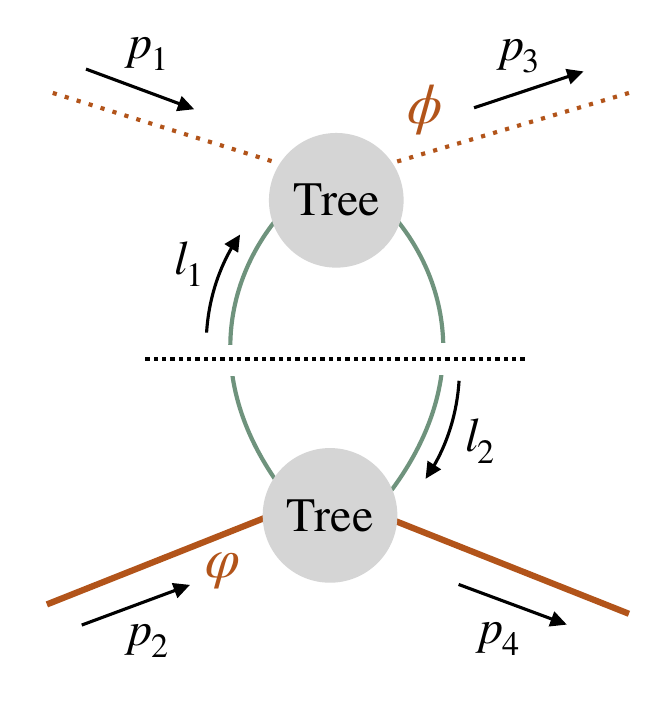}
    \caption{Scattering of one massless scalar (dotted orange line) and one massive scalar (solid
orange line) mediated by a graviton (solid blue line) at $\mathcal{O}(G^2)$. The dotted black line represents the discontinuity cut.}
    \label{fig:t_cut}
\end{figure}

We consider $2\rightarrow 2$ scattering of a massless scalar $\phi$ interacting gravitationally with a heavy massive scalar $\varphi$. The tree-level graviton exchange in the $M\gg\omega\sim |\mathbf{q}|$ limit is simply given by
\begin{equation}
    T^{(0)}(\mathbf{p},\mathbf{p}^\prime)=\frac{32\pi GM^2\omega^2}{\mathbf{q}^2}\,.
\end{equation}
For the one-loop calculation, we restrict ourselves to non-analyticities in the transferred momentum $\mathbf{q}$, assuming that no contact terms contribute at this order\footnote{Contact terms become necessary when UV divergences first appear, which are known to occur at $\mathcal{O}(G^3)$, see \cite{Ivanov:2024sds}.}. These give rise to long distance effects and are picked up by the Fourier transform of section \ref{sec:Todorov}. These effects are fully captured by considering the $t$-channel cut represented in Figure \ref{fig:t_cut}, on which unitarity dictates that the amplitude factorizes into two tree-level amplitudes as 
\begin{align}\label{eq:tcut}
  i  T^{(1)}(\mathbf{p},\mathbf{p}^\prime)\bigg|_{\text{cut-t}}&=\frac{\mu^{2\epsilon}}{2}\int \d\text{LIPS}(l_1,-l_2)\notag\\
  &\sum_{\lambda_1,\lambda_2}\mathcal{M}(p_1,l_1^{\lambda_1},-p_3,-l_2^{\lambda_2}) \times \mathcal{M}(p_2,-l_1^{\lambda_1},-p_4,l_2^{\lambda_2})^\dagger
\end{align}
where the sum runs over all possible helicity configurations of the internal gravitons, the Lorentz invariant phase-space is  $d\text{LIPS}(l_1,-l_2)=d^4l_1 d^4l_2\delta^{(+)}(l_1^2)\delta^{(+)}(l_2^2)\delta^4(p_1+p_3+l_1-l_2)$, and  $\mathcal{M}(i,k^{\lambda_1},j,l^{\lambda_2}) $ is the tree-level 4-point amplitude with all incoming momenta between two scalars with momenta $i$ and $j$ and two gravitons with momenta $k$ and $l$ and helicity of $\lambda_1$ and $\lambda_2$ respectively (see Figure \ref{fig:t_cut}).\footnote{Here we make use of standard notation in the on-shell approach to computing Feynman diagrams (for reviews see e.g. \cite{Cheung:2017pzi,Elvang:2013cua}).} 

Since the tree-level 4-point amplitudes for the massless state $\phi$ vanishes for helicity flipping gravitons, i.e. $\mathcal{M}(1_\phi,2^+,3_\phi, 4^+)=\mathcal{M}(1_\phi,2^-,3_\phi, 4^-)=0$ \cite{Bjerrum-Bohr:2017dxw,Bjerrum-Bohr:2016hpa}, we only need to consider the helicity preserving configuration, where the relevant amplitudes are
\begin{align}\label{eq:Mallin}
    \mathcal{M}(1_\phi,2^-,3_\phi, 4^+)&=8\pi G\frac{\langle 2 1 4]^4}{s(t-M^2)(u-M^2)}\, ,\notag\\ 
   \mathcal{M}(1_\varphi,2^-,3_\varphi, 4^+)&=8\pi G\frac{\langle 2 1 4]^4}{stu}\, ,
\end{align}
here given for all in-going external legs.
We replace the amplitudes \eqref{eq:Mallin} into (\ref{eq:tcut}), and compute explicitly the numerator by tracing over Pauli matrices. Then, we employ LiteRed \cite{Lee:2012cn,Lee:2013mka} to project the result into a basis of one-loop master integrals (MIs). Lastly, we impose the Born regime $M\gg \omega\sim \mathbf{q}$ on the obtained coefficients, finding that the amplitude at leading order takes the form
 \begin{equation}\label{eq:oneloopIBP}\notag
i\, T^{(1)}(\mathbf{p},\mathbf{p}^\prime)=c_{\Box}\mathcal{I}_\Box+c_{\mathrm{X}}\mathcal{I}_\mathrm{X}+c_{\triangle}\mathcal{I}_\triangle\, ,
\end{equation}
where $\mathcal{I}_\Box$, $\mathcal{I}_{\mathrm{X}}$ and $\mathcal{I}_\triangle$ are the master integrals respectively of box, cross-box and triangle. Notice that the massless (opposite side) triangle does not contribute to the wave regime, since it can be projected into a bubble master integral. This feature is also present in the eikonal computation \cite{Bjerrum-Bohr:2017dxw,Bjerrum-Bohr:2016hpa}.

At leading order in the heavy mass limit, the coefficients are given by
\begin{align}\label{eq:coeffIBP}
    c_{\Box}=c_{\mathrm{X}}=1024\pi^2 G^2 \omega^4 M^4\, ,\notag\\
    c_\triangle=-16 \pi^2G^2M^4(60\, \omega^2-\mathbf{q}^2)\, .
\end{align}
First of all, let us focus on the box and cross-box pieces, whose contributions combine to reproduce precisely the iteration term $\sim V G V$, as shown explicitly in Appendix \ref{app:iteration}. This observation is highly reminiscent of the computation in the eikonal regime, where box and cross-box give the superclassical piece that leads to the `square' of the eikonal phase at tree-level.\footnote{Despite this similarity, we want to stress that in angular momentum space, the iteration of the tree-level potential in the Born regime is given by a digamma $\sim \big[G \psi^{(0)}(J-1)\big]^2$ which in the eikonal limit $J \to \infty$ gives precisely the $( G \log J)^2$ iteration.} 

After having identified the iteration piece, we focus on the remaining contribution from the triangle topology. We can identify two effects, one that dominates in the eikonal $\omega\gg \mathbf{q}$ limit and a second piece. Since in the large mass limit, the triangle master integral is given by 
\begin{equation}
    \mathcal{I}_\triangle=-\frac{i}{32M |\mathbf{q}|}+\mathcal{O}(1/M^2)\,,
\end{equation}
we immediately identify the first piece as the usual 2PM correction \cite{Bjerrum-Bohr:2016hpa}, that enters in the bending of light problem and allows one to extract the classical scattering angle $\theta=\frac{4GM}{b}+\frac{15\pi G^2M^2}{4b^2}$. We will identify the origin of the second contribution in section \ref{sec:RW} by matching directly to GR.

In conclusion, we can merge all the results, recovering that up to $\mathcal{O}(G^2)$ the amplitude takes the following Born series form
\begin{align}\label{eq:2PM_tot}
    T(\mathbf{p},\mathbf{p}^\prime)=\left(V^{(0)}(\mathbf{p},\mathbf{p}^\prime) + V^{(1)} (\mathbf{p},\mathbf{p}^\prime)\right)+ \int \d^3\mathbf{k}\, V^{(0)}(\mathbf{p},\mathbf{k})G(\mathbf{p},\mathbf{k}) V^{(0)}(\mathbf{k},\mathbf{p}^\prime)\, ,
\end{align}
with 
\begin{align}\label{eq:V01pm}
    V^{(0)}(\mathbf{p},\mathbf{p}^\prime)&=\frac{32\pi GM^2\omega^2}{\mathbf{q}^2}\,\notag ,\\
    V^{(1)}(\mathbf{p},\mathbf{p}^\prime)&=\frac{30\pi^2 G^2M^3\omega^2}{|\mathbf{q}|}-\frac{G^2M^3\pi^2}{2}|\mathbf{q}|\, .
\end{align}
which matches precisely the expected structure of a Born series, where $V(\mathbf{p},\mathbf{p}^\prime)$ only has singularities for $\mathbf{q} \to 0$ corresponding to the long-range gravitational interaction, as discussed in the previous Chapter. In the next section we recover this result from the Regge-Wheeler equation.

\subsection{The Regge-Wheeler equation}\label{sec:RW}
In Chapter \ref{sec:Todorov} we observed that in the Born regime of one massless and one massive external states, the amplitude reduces to a Born series solution of a classical wave equation. Since we are studying a massless wave propagating in a Schwarzschild background, it is expected that the potential (\ref{eq:V01pm}) should emerge from the corresponding classical wave equation, which is the Regge-Wheeler equation  \cite{PhysRev.108.1063}. In the following, we extract the potential at $\mathcal{O}(G^2)$ from the Regge-Wheeler equation and explicitly match it to the on-shell potential computed previously in (\ref{eq:V01pm}).

We consider a minimally coupled massless scalar field on a generic gravitational background
\begin{equation}
S = \int \d^4 x \sqrt{-g} \, g^{\mu \nu} \partial_\mu \phi \,\partial_\nu \phi \, ,
\end{equation}
whose wave equation is simply recovered from the variational principle 
\begin{equation}
{\delta S \over \delta \phi} = 0 \qquad \implies \qquad \partial_\mu(\sqrt{-g} \, g^{\mu \nu} \partial_\nu \phi) = 0.
\end{equation}
We are interested in evaluating this equation of motion on a Schwarzschild metric. Here we work in isotropic cartesian coordinates\footnote{This is a convenient but not necessary choice of coordinate system. In Appendix \ref{app:schw_coord} we exemplify how different choices lead to the same potential, once on-shell conditions are imposed. }, for which the metric reads
\begin{equation}
    g_{\mu \nu} = \begin{pmatrix}
A(r) &0 & 0 & 0\\
0 & -B(r) & 0&0\\
0&0& -B(r) & 0\\
0 & 0 & 0&- B(r)
\end{pmatrix}\,, \qquad A(r) = {\big( 1 - {G M \over 2 r}\big)^2 \over \big( 1 + {G M \over 2 r}\big)^2 }\, , \quad B(r) = \Big( 1 + {G M \over 2 r}\Big)^4\, ,
\end{equation}
where we used the standard notation where $G$ is the Newton's constant, $M$ is the black hole mass, and $r=x^2+y^2+z^2$.
Evaluated on this metric the wave equation takes the following form
\begin{equation}
      \nabla^2 \phi(x) = \left({B \over A} \partial_t^2 - {\partial_r (AB) \over 2 A B} \partial_r  \right)\phi(x)
\end{equation}
with $\nabla^2 = \delta^{ i j} \partial_i \partial_j$ the Euclidean Laplacian and $\partial_r= {1 \over r} \, \mathbf{x} \cdot \nabla$. This equation is the Regge-Wheeler equation. Since the problem is time independent and spherically symmetric, we have $\phi(x)=e^{-i\omega t}\phi(\mathbf{x})$, and the above becomes
\begin{equation}
    (\nabla^2+\omega^2)\phi(\mathbf{x})=\left(\left(1-\frac{B}{A}\right)\omega^2- {\partial_r (AB) \over 2 A B} \partial_r  \right)\phi(\mathbf{x})
\end{equation}

By matching to equation \eqref{eq:cweq}, we can easily extract the potential $V$ at all orders in $G$
\begin{equation}\label{eq:allordV}
 V =  \left(1-\frac{\left(1+\frac{ G M}{2r}\right)^6}{\left(1-\frac{ G M}{2r}\right)^2} \right)\omega^2  + \frac{2G^2M^2}{G^2M^2r-4r^3}\partial_r\, ,
\end{equation}
which at this level is a differential operator acting on the scalar wave $\phi$. 
 The potential up to $\mathcal{O}(G^2)$ becomes
\begin{equation}\label{eq:V2PM}
  V =  -\left(\frac{4 MG}{r}+\frac{15G^2 M^2 }{2 r^2} \right) \omega^2 - \frac{G^2 M^2}{2 r^3} \partial_r  + \mathcal{O}(G^3)\, .
\end{equation}
Notice that the first bracket dominates at high frequencies. Indeed, this is the only term that survives in the eikonal (or WKB limit) \cite{Cheung:2018wkq} and contributes to eikonal observables such as the scattering angle. 

The additional term does not scale with the frequency. In order to identify how this contribution affects the amplitude, we Fourier transform the potential to momentum space following \eqref{eq:VEOB}
\begin{align}\label{eq:potentialq}
    V(\mathbf{p},\mathbf{p}^\prime)&=-2M \int \d^3\mathbf{x}\,  e^{-i\mathbf{p}'\cdot \mathbf{x}}V(r) e^{i\mathbf{p}\cdot  \mathbf{x}}\notag\\
    &=\frac{32\pi G M^2\omega^2}{\mathbf{q}^2}+\frac{30\pi^2 G^2M^3\omega^2}{|\mathbf{q}|}-\frac{G^2M^3\pi^2}{2}|\mathbf{q}|
\end{align}
where we used the Fourier transforms relations of Appendix \ref{app:FT} and imposed the on-shell relation $\mathbf{p}\cdot \mathbf{p}^\prime=\mathbf{q}^2/2$. 
Once again we recognise the analytic structure in $\mathbf{q}$ coming from the graviton tree-level exchange at $\mathcal{O}(G)$ and from the triangle master integral $\mathcal{O}(G^2)$. This potential exactly matches the amplitude result (\ref{eq:V01pm}). In Appendix \ref{app:schw_coord}, we show that the same result is obtained by considering a different choice of coordinate system (Schwarzschild coordinates), which interestingly leads to a different off-shell potential. It nonetheless matches (\ref{eq:potentialq}) when on-shell conditions are imposed on the external legs. 

It is easy to see that subleading corrections to the $M\gg \omega\sim |\mathbf{q}|$ limit (which in the massive case are given by the gravitational self-force expansion) are always quantum corrections, as the only dimensionless small parameters of this expansion are $\omega/M$ and $|\mathbf{q}|/M$. 

In the following Chapter we will consider massless propagations on a Kerr background instead.

\section{Kerr-Compton as the Born amplitude}\label{sec:Kerr}
In this section, we exploit the Born regime discussed in section \ref{sec:Todorov}, to compute amplitudes for massless waves propagating on a Kerr background. We focus on scalar and spin-1 massless wave at leading order in $\mathcal{O}(G)$ and any order in the spin parameter $a^\mu$. These amplitudes have been object of intensive study in the last years, mostly from a theoretical perspective of recovering classical spin dynamics from the large spin limit of quantum scattering amplitudes (see for instance \cite{Porto:2005ac,Levi:2015msa,Saketh:2022wap,Jakobsen:2023ndj,Ben-Shahar:2023djm,Arkani-Hamed:2019ymq,Guevara:2018wpp,Chung:2019duq,Aoude:2020onz,Chung:2018kqs,Arkani-Hamed:2017jhn,Bautista:2021wfy,Bautista:2022wjf,Bautista:2023sdf,Scheopner:2023rzp,Vines:2017hyw,Kosmopoulos:2021zoq,Chen:2021kxt,Aoude:2022trd,Bern:2022kto,Aoude:2022thd,FebresCordero:2022jts,Menezes:2022tcs,Bjerrum-Bohr:2023jau,Aoude:2023vdk,Bern:2023ity,Bjerrum-Bohr:2023iey,Chen:2023qzo,Luna:2023uwd,Guevara:2019fsj,Cangemi:2022bew,Cangemi:2022abk,Cangemi:2023ysz,Chia:2020yla,Cangemi:2023bpe}). In Chapter \ref{sec:2PM} we reproduced the Schwarzschild background at 2PM, by considering a minimally coupled scalar, and we expect that an analogous story should hold for a Kerr geometry, emerging from some minimally coupled higher spin particle. To date, the most reliable way to identify correctly which amplitude describes a Kerr black hole interaction is via the matching to the BHPT results \cite{Mano:1996vt,Mano:1996gn,Mano:1996mf,Sasaki:2003xr,Bonelli:2022ten,Dodelson:2022yvn,Aminov:2023jve,Bautista:2023sdf}. The answer that comes out of BHPT is arguably hard to manipulate, because it takes place in angular momentum space in terms of partial waves. A challenge of this approach is to isolate contact terms related to conservative dynamics and those emerging from the boundary conditions at the horizon. As far as we understand, this challenge is not resolved, resulting in multiple prescriptions on how to extract conservative pieces. 

In the following, we present an alternative way to match Kerr amplitudes directly to the wave equation, exploiting the understanding of the Born regime of section \ref{sec:Todorov}. Our approach presents multiple advantages. First of all, at leading order in $G$, the Born series does not include any iteration, and therefore there is a simple identification $T=V^{(0)}+ \mathcal{O}(G^2)$. Hence, the result is achieved by simply extracting the potential from the equation of motion and then Fourier transform it. This means that we work only with the leading order in $G$ from the very beginning, without carrying around an all-order solution, which includes a lot of unnecessary information. Another advantage is that since the PM expansion is performed at the level of the wave equation, all dissipative effects are immediately neglected, and we can directly target only conservative contact terms, which are independent on the choice of boundary conditions (the expansion in $G$ erases the presence of an horizon). Lastly, the potential has a simple analytic structure in the spin parameter $a^\mu$ and can be expanded in $a$ without any subtleties, on the contrary of \cite{Bautista:2021wfy,Bautista:2022wjf}, where a non-unique analytic continuation into the super-extremal regime needs to be performed on the partial wave solution.

We find it convenient to use the Kerr-Schild coordinate system which allows us to define a Lorentz covariant potential. The metric takes the form
\begin{equation}\label{eq:KSmetric}
    g_{\mu\nu}=\eta_{\mu\nu}-\Phi l_\mu l_\nu \, , \qquad g^{\mu\nu}=\eta^{\mu\nu}+\Phi l^\mu l^\nu \, ,
\end{equation}
and we work with a covariant spin vector $a_i$ such that
\begin{align}\label{eq:defPhil}
    \Phi&=\frac{2GM r^3}{r^4+(\mathbf{a}\cdot \mathbf{x})^2}\, \notag ,\\
    l_\mu& =\left( 1\, , \,\frac{1}{r^2+a^2}\left[ r x_i+\frac{(\mathbf{a}\cdot \mathbf{x})}{r}a_i+ \epsilon_{ijk} x_j a_k\right]\right)\, ,
\end{align}
where $a^2=a_i a_i$ and the obloid radius $r$ satisfies the following relation
\begin{equation}
    a^2+r^2=|\mathbf{x}|^2 +\frac{(\mathbf{a}\cdot \mathbf{x})^2}{r^2}\, .
\end{equation}
Notice that the Schwarzschild metric is recovered by simply setting $a_i=0$.

\subsection{Scalar wave}
We consider a massless scalar minimally coupled to a Kerr metric in Kerr-Schild coordinates (\ref{eq:KSmetric}). The kinetic term is given by
\begin{align}\label{eq:lagrscaKS}
    \mathcal{L}=\nabla^\mu \phi(x)\nabla_\nu \phi(x)=\partial^\mu\phi(x)\partial_\mu\phi(x)+\Phi l^\mu l^\nu\partial_\mu \phi(x)\partial_\nu \phi(x)\, ,
\end{align}
and the equations of motion take the following form
\begin{align}
  &  \partial_\mu\partial^\mu \phi(x)+\partial_\mu\left(\Phi\,  l^\mu l^\nu \partial_\nu \phi(x)\right)=0\,,
\end{align}
which can be recast as 
\begin{equation}
    (\nabla^2+\omega^2)\phi(\mathbf{x})=\left[-\Phi \omega^2 + i\omega\left[ \partial_i(\Phi l_i)+2\Phi l_i \partial_i\right]+\partial_i(\Phi l_il_j\partial_j)\right]\phi(\mathbf{x})
\end{equation}
by imposing $\phi(x)=e^{-i\omega t}\phi(\mathbf{x})$.
By matching to \eqref{eq:cweq}, we identify the potential 
\begin{align}\label{eq:VscnonXd}
    V(\mathbf{x})=&-\Phi \omega^2 +i\omega\big[ \partial_i(\Phi l_i)+2\Phi l_i \partial_i\big]+\partial_i(\Phi l_il_j\partial_j)\notag\\
    =&-\frac{2GM \omega^2 r^3}{r^4+(\mathbf{a}\cdot \mathbf{x})^2} \notag\\
    &+\frac{2i GM \omega r^2}{r^4+(\mathbf{a}\cdot \mathbf{x})^2}\bigg[ 1+\frac{2\left((\mathbf{a}\cdot \mathbf{x})a_i+r^2 x_i\right)\partial_i}{r^2+a^2}-\frac{2 r\epsilon_{ijk}a_jx_k\partial_i}{r^2+a^2}\bigg] \notag\\
    &+\frac{2GMr}{(r^2+a^2)(r^4+(\mathbf{a}\cdot \mathbf{x})^2)}\Big[\big((\mathbf{a}\cdot \mathbf{x})a_i+r^2 x_i\big)\partial_i -r\epsilon_{ijk}a_jx_k\partial_i\Big]\notag \\
    &-\frac{2GM r^3}{(r^2+a^2)^2(r^4+(\mathbf{a}\cdot \mathbf{x})^2)}\bigg[\Big(r^2+ a^2-\frac{2(\mathbf{a}\cdot \mathbf{x})^2}{r^2}\Big) a_ia_j\partial_i\partial_j  \notag\\
    &+(a^2-r^2) x_ix_j\partial_i\partial_j-(r^2+a^2)\Big( a^2-\frac{(\mathbf{a}\cdot \mathbf{x})^2}{r^2}\Big)\partial_i\partial_i  \notag\\
    &-4(\mathbf{a}\cdot \mathbf{x})a_ix_j\partial_i\partial_j+2\Big(r x_i\partial_i +\frac{(\mathbf{a}\cdot \mathbf{x})}{r}a_i\partial_i\Big)\epsilon_{jkl}a_k x_l \partial_j\bigg]\,,
\end{align}
where we replaced the metric (\ref{eq:defPhil}) explicitly.

We perform the Fourier transform of this expression by Taylor expanding the integrand order by order in $a_i$ and applying the integral and combinatorial identities in Appendix \ref{app:FT}. The result in momentum space beautifully resums the spin dependence into simple exponentials. Since the amplitude at $\mathcal{O}(G)$ in the Born regime reduces to $T=V$, we can identify our result with the amplitude itself, 
\begin{align}\label{eq:VscnonX}
   T(\mathbf{p},\mathbf{p}^\prime) &= \frac{32 \pi  G M^2}{\mathbf{q}^2} \Big[ \, \omega ^2 \cosh (\mathbf{a} \cdot \mathbf{q})-i \omega \epsilon_{i j l} q_i a_j p_l \frac{\sinh (\mathbf{a} \cdot \mathbf{q})}{(\mathbf{a} \cdot \mathbf{q})}\Big]  \\
    &\qquad + 8 \pi G M \big[ a^2 \mathbf{q}^2 - 2 (\mathbf{a} \cdot \mathbf{q}) (\mathbf{a} \cdot \mathbf{p}^\prime) \big] { (\mathbf{a} \cdot \mathbf{q}) \cosh(\mathbf{a} \cdot \mathbf{q}) - \sinh({\mathbf{a} \cdot \mathbf{q}}) \over (\mathbf{a} \cdot \mathbf{q})^3}\, .\notag
\end{align}
A quick inspection of (\ref{eq:VscnonX}) reveals that this result is not crossing symmetric (transformation $\mathbf{p}\leftrightarrow -\mathbf{p}^\prime$ and $\omega\rightarrow -\omega$). This is clearly a violation of a fundamental principle of the S-matrix. It turns out that in our derivation we are missing some surprising contributions from the boundary at infinity.

To understand their origin, we re-derive the Euler-Lagrange equations for the Lagrangian (\ref{eq:lagrscaKS}), by applying the variational principle
\begin{align}\label{eq:EOMwBT}
-  \partial_\mu\partial^\mu \phi(x)=\partial_\mu\left(\Phi\,  l^\mu l^\nu \partial_\nu \phi(x)\right) -\int \d^4y \, \partial_\mu\left[\delta^4(x-y) \left( \partial^\mu+\Phi l^\mu l^\nu \partial_\nu\right) \phi(y)\right]\, ,
\end{align}
where we used the fact that in Kerr-Schild coordinates the determinant satisfies $g=1$. The total derivative contribution appears from the integration by part of the derivative acting on a delta function. 
This piece, which can be rewritten as boundary term at infinity, is usually set to $0$ in most common instances. It turns out that in our case this contribution does not vanish, and on the contrary contributes to the amplitude and restores crossing symmetry. Despite it being an unusual term in amplitude calculations, it can be computed explicitly and arises at $\mathcal{O}(a^2)$. 

For concreteness, let us show an explicit example. Replacing $\Phi$ and $l^\mu$ with their definition (\ref{eq:defPhil}) in the second line of (\ref{eq:EOMwBT}) and keeping the piece proportional to $x_i x_j$ at order $\mathcal{O}(a^2)$, the total derivative contribution becomes 
\begin{align}\label{eq:boundexample}
  a^2 & \int \d^3 \mathbf{x}\,  \partial_i\left(\frac{x_i x_j}{r^5} e^{-i\mathbf{p}'\cdot\mathbf{x}}\partial_j (e^{i\mathbf{p}\cdot \mathbf{x}})\right)= ia^2p_j\int_{S^2} \d^2x \, r^2 \, n_i \frac{x_i x_j}{r^5}e^{- i \mathbf{q}\cdot \mathbf{x}}\notag \\
    &= a^2 p_j q_i\int \d^2x \frac{x_i x_j}{r}=\frac{4\pi a^2}{3}\mathbf{p}\cdot\mathbf{q}\, ,
\end{align}
where in the first line we applied Stokes theorem, expanded the exponential in an all order Taylor expansion in the second line and noticed that all except the linear term vanishes as $r \to \infty$. This is clearly a finite effect that contributes on-shell. By explicit inspection, we notice that all total derivative contributions of this type lead to contact terms, and therefore do not impact in any way observables such as the scattering angle in the eikonal limit. They are however important in the Born regime.

We can incorporate the boundary terms in a manifestly crossing symmetric form for the potential without having to compute explicitly a boundary integral as done above.
In particular, by applying $ -16\pi^3M\int d^4x e^{ip_{3} \cdot x}$ on both sides of (\ref{eq:EOMwBT}), and replacing $\phi(x)=\int\frac{\d^4 k}{(2\pi)^4}\phi(k)e^{-ik\cdot x}$ the wave equation becomes
\begin{align}
    \phi(\mathbf{p}^\prime)&= G(\mathbf{p}^\prime,\omega)\left[16\pi^3M \int \d^4x \,\partial_\mu( e^{ip_3\cdot x})\, \Phi l^\mu l^\nu  \partial_\nu \phi(x)\right]\notag\\
    &=G(\mathbf{p}^\prime,\omega)\int \d^3\mathbf{k} \, \phi(\mathbf{k})\left[2M p_{3\mu} k_\nu \int \d^3\mathbf{x}\, \Phi l^\mu l^\nu e^{i(\mathbf{k}-\mathbf{p}^\prime)\cdot\mathbf{x}}\right]\, ,
\end{align}
where we commuted the integral over $x$ inside the derivative (which is with respect to $y$) and neglected analytic contributions. The combination of the two terms proportional to $G$ (or equivalently to $\Phi$) combine in such a way that the partial derivative acts on the phase $e^{ip_3\cdot x}$. Notice that the integral over the time component simply leads to the condition $k_0=\omega$.

By matching to the wave equation in momentum space \eqref{eq:eobms}, we extract the potential 
\begin{equation}\label{eq:Vscll}
 V_{\phi}(\mathbf{p},\mathbf{p}^\prime)  = 2M \, p_{3\mu}\,  p_{1\nu}  \int \d^3\mathbf{x} \,  \Phi l^\mu l^\nu e^{i(\mathbf{p}-\mathbf{p}^\prime)\cdot\mathbf{x}}\, .
\end{equation}
where $p_{1\mu} = (\omega, \mathbf{p})$ and $p_{3\mu} = (\omega, \mathbf{p}')$ are respectively the 4-momenta of the incoming and outgoing massless scalar. Notice that the additional contribution from the boundary term precisely combines with (\ref{eq:VscnonXd}) to produce a crossing symmetric potential.

We now project the Fourier transform in (\ref{eq:Vscll}) using the covariant projectors 
\begin{equation}\label{eq:projdef}
    \Pi_T^{\mu\nu}=\eta^{\mu\nu}-u^\mu u^\nu\,, \quad \Pi_L^{\mu\nu}=u^\mu u^\nu\, , \quad u_\mu=(1,0,0,0)\, ,
\end{equation}
where $Mu^\mu$ is the 4-momentum of the black hole. We replace $l^\mu=u^\mu l_0+\Pi_T^{\mu i}l_i$, obtaining
\begin{align}\label{eq:projll}
    \int \d^3 \mathbf{x} \, \Phi l^\mu l^\nu e^{i(\mathbf{p}-\mathbf{p}^\prime)\cdot\mathbf{x}}&=u^\mu u^\nu \int \d^3 \mathbf{x} \,\Phi  \, e^{i(\mathbf{p}-\mathbf{p}^\prime)\cdot\mathbf{x}}+u^\mu \Pi_T^{\nu j}\int \d^3 \mathbf{x} \,\Phi \,l_j\, e^{i(\mathbf{p}-\mathbf{p}^\prime)\cdot\mathbf{x}}\notag \\
   & +\Pi_T^{\mu i}u^\nu \int \d^3 \mathbf{x}\, \Phi \,l_i \,e^{i(\mathbf{p}-\mathbf{p}^\prime)\cdot\mathbf{x}}+ \Pi_T^{\mu i}\Pi_T^{\nu j}\int \d^3 \mathbf{x} \,\Phi\, l_i l_j e^{i(\mathbf{p}-\mathbf{p}^\prime)\cdot\mathbf{x}}
\end{align}
where we used $l_0=1$.

Now, we replace the explicit Kerr-Schild metric \eqref{eq:KSmetric}, and perform the Fourier transform by making use of the identities in Appendix \ref{app:FT}. This results in multiple tensor structures canceling when contracted and the remaining pieces resuming into exponentials.\footnote{Note in particular that many of these Fourier transforms are IR divergent by themselves, see equation \eqref{eq:Vmq}. They have to combine in the right way for the IR divergence contributions to cancel in the potential. This provides a non-trivial check of the derivation.}

In conclusion, from (\ref{eq:Vscll}) we obtain the full $\mathcal{O}(G)$ conservative scalar potential, and therefore amplitude 
\begin{align}\label{eq:VscX}
 T_{\phi}(\mathbf{p},\mathbf{p}^\prime) &= \frac{32 \pi  G M^2}{\mathbf{q}^2} \Big[ \, \omega ^2 \cosh (\mathbf{a} \cdot \mathbf{q})-i \omega \epsilon_{i j l} q_i a_j p_l \frac{\sinh (\mathbf{a} \cdot \mathbf{q})}{(\mathbf{a} \cdot \mathbf{q})}\Big] \\
    &\qquad + 16 \pi G M^2 \big[(\mathbf{a} \cdot \mathbf{p}^\prime) (\mathbf{a} \cdot \mathbf{p})- a^2 \omega^2  \big] { (\mathbf{a} \cdot \mathbf{q}) \cosh(\mathbf{a} \cdot \mathbf{q}) - \sinh({\mathbf{a} \cdot \mathbf{q}}) \over (\mathbf{a} \cdot \mathbf{q})^3}\, ,\notag
\end{align}
or, alternatively, in fully covariant form
\begin{align}\label{eq:VscXCov}
     T_{\phi}(p_1,p_3) &= -\frac{32 \pi  G M^2}{q^2}\bigg[ \, (u\cdot p_1) (u\cdot p_3) \cosh (a\cdot q)\\&+{i \over 2} \left({u\cdot p_1+u\cdot p_3}\right) \epsilon_{\alpha\beta\gamma\delta} a^{\alpha}p_3^\beta p_1^\gamma u^\delta \,\frac{\sinh (a\cdot q)}{(a\cdot q)}\bigg]\notag\\
     &+16 \pi G M^2 \big[(a \cdot p_1)(a \cdot p_3) - a^2 (u\cdot p_1) (u\cdot p_3)  \big]{ (a\cdot q)\cosh(a \cdot q) - \sinh(a \cdot q) \over (a \cdot q)^3}\, ,\notag
\end{align}
where we recall $p_{2\mu}=p_{4\mu}=Mu_{\mu}$, as expected in the probe limit, and $a^\mu$ and $q^\mu$ are orthogonal to $u^\mu$, $a.u = 0$ and $q.u = 0$. We checked the resummation up to $\mathcal{O}(a^{30})$.

The result (\ref{eq:VscXCov}) showcases a number of features. First of all, the first two lines of (\ref{eq:VscXCov}) contribute to the $t$-channel residue and match a known result obtained by gluing 3-point functions as dictated by factorization\,\footnote{The 3-point function of a Kerr black hole and one graviton can be extracted by Fourier transforming its stress-energy tensor $T_{\mu\nu}$, see \cite{Vines:2017hyw,Harte:2016vwo}.} (see e.g. \cite{Bautista:2021wfy}). Secondly, the last piece of (\ref{eq:VscXCov}) is a contact term despite not looking like it at first glance. It does not contribute to the $t$ pole due to the interplay between the $\cosh(a\cdot q)$ and $\sinh{(a\cdot q)}$ terms making this contribution vanish as $|\mathbf{q}|\rightarrow 0$. Moreover, the amplitude (\ref{eq:VscXCov}) is crossing symmetric, as opposed to \eqref{eq:VscnonX}, highlighting once again the importance of the boundary terms emerging from the total derivative of \eqref{eq:EOMwBT}. Finally, the scalar amplitude in the eikonal limit $\omega\gg |\mathbf{q}|$ takes the form 
\begin{equation}\label{eq:scEik}
   T_{\phi}(p_1,p_3)\xrightarrow[M\gg \omega\gg |\mathbf{q}|]{} \frac{32 \pi  G M^2}{\mathbf{q}^2} \Big[ \, \omega ^2 \cosh (\mathbf{a} \cdot \mathbf{q})-i \omega \epsilon_{i j l} q_i a_j p_l \frac{\sinh (\mathbf{a} \cdot \mathbf{q})}{(\mathbf{a} \cdot \mathbf{q})}\Big]\, ,
\end{equation}
which has been shown to lead to the correct classical deflection angle for a null geodesic on a Kerr background \cite{Arkani-Hamed:2019ymq}.

The fact that we obtained some contact pieces (last line of (\ref{eq:VscXCov})) from our derivation might come as a surprise. Since the potential is the result of a Fourier transform of a long distance potential, we would naively expect that local contributions would not have support when integrated against a wavepacket, as they are conjugate to delta functions in the potential $V(\mathbf{x})$. These contact terms are actually non-analytic pieces when first Fourier transformed and reduce to contact terms only when on-shell conditions are imposed. Similarly to the discussion in Chapter \ref{sec:2PM}, we expect that different off-shell continuations will lead to the Kerr potential in different coordinate systems.

As far as we can tell this result is not present in the literature and we were not able to match it to \cite{Bautista:2021wfy}. We expect that the reason lies in the fact that the derived contact terms mix with dissipation/horizon effects, which contribute in a similar manner in the Born regime. It would be interesting to check how our result affects partial waves and combines with dissipation effects, hopefully shedding some light on the origin of different contributions in BHPT. We leave this avenue for future work. 

\subsection{Photon}
After the derivation of the Kerr amplitude for a scalar wave, we are ready to address a similar calculation of a massless spin-1 propagating on a Kerr background. The result is obtained in a very analogous way. 

We consider the Lagrangian of a massless spin-1 on a Kerr background expressed in Kerr-Schild coordinates \eqref{eq:KSmetric}
\begin{align}\label{eq:phoL}
    \mathcal{L}=&-\frac{1}{4}F_{\mu\nu}F^{\mu\nu}=-\frac{1}{4}g^{\mu\alpha}g^{\nu\beta}F_{\mu\nu}F_{\alpha\beta}\notag\\
    =&-\frac{1}{4}\eta^{\mu\alpha}\eta^{\nu\beta}F_{\mu\nu}F_{\alpha\beta}-\frac{1}{2}\Phi l^\mu l^\alpha \eta^{\nu\beta} F_{\mu\nu}F_{\alpha\beta}+ \mathcal{O}(\Phi^2)\,,
\end{align}
where $F_{\mu\nu}=\nabla_\mu A_\nu-\nabla_\nu A_\mu=\partial_\mu A_\nu-\partial_\nu A_\mu$ and we neglect terms in $\Phi^2$ (defined in \eqref{eq:defPhil}), as we are only interested in the leading piece in $G$.

 The equations of motion are derived by applying the variational principle to \eqref{eq:phoL}, and are given by
\begin{align}\label{eq:phoEOM}
    -\eta^{\mu\alpha}\eta^{\nu\beta}\partial_\mu& F_{\alpha\beta} 
    \notag=\partial_\mu \left[ \Phi \left(\eta^{\nu\beta}   l^\mu l^\alpha + \eta^{\alpha\mu} l^\beta l^\nu \right) F_{\alpha\beta}\right]\notag\\
    &- \int \d^4y\partial_\mu\left[ \delta^4(x-y)\Phi \left(\eta^{\nu\beta}   l^\mu l^\alpha + \eta^{\alpha\mu} l^\beta l^\nu \right) F_{\alpha\beta}\right]+\mathcal{O}(\Phi^2)\,.
\end{align}

Analogously to the scalar derivation, we apply $ -16\pi^3M\int d^4x e^{ip_{3} \cdot x}$ on both sides of \eqref{eq:phoEOM}, obtaining 
\begin{align}\label{eq:phoEOMp}
    &A_\nu(\mathbf{p}^\prime)=G(\mathbf{p}^\prime,\omega)\left[16 \pi^3 M\int \d^4x\, \partial_\mu\left(e^{ip_{3} \cdot x}\right)\Phi \left(\eta^{\nu\beta}   l^\mu l^\alpha + \eta^{\alpha\mu} l^\beta l^\nu \right) F_{\alpha\beta}(x)\right]\\
 &   \!\!=G(\mathbf{p}^\prime,\omega)\int \d^3\mathbf{k} \,A_\mu(\mathbf{k})\left[2M \left( k_\alpha \delta^\mu_\beta -k_\beta \delta^\mu_\alpha \right)\left( p_{3\sigma}\eta^{ \nu\beta}\delta^\alpha_\rho + p_3^\alpha \delta_\sigma^\beta \delta_\rho^\nu\right)\int \d^3 \mathbf{x}\, \Phi\, l^\sigma l^\rho  e^{i(\mathbf{k}-\mathbf{p}^\prime)\cdot \mathbf{x}}\right]\notag
\end{align}
where we imposed Lorentz gauge on the flat metric $\eta^{\mu\nu}\partial_\mu A_\nu =0$. Notice that also in this case, the two terms on the l.h.s of \eqref{eq:phoEOM} combine such that the derivative $\partial_\mu$ acts on the phase $ e^{ip_{3}\cdot x}$. 

In this work, we do not provide a full derivation of the Lippmann-Schwinger equation in the Born regime for a spin-1 particle (as opposed to the scalar results in Chapter \ref{sec:EOB} and Appendix \ref{app:LS}), but nonetheless we extract a potential from \eqref{eq:phoEOMp}, under the assumption that an equivalent computation will hold. 

In this case, the potential is given by a tensor structure with two free indices 
\begin{align}\label{eq:Vphoton}
    V_\gamma^{\mu\nu}(\mathbf{p},\mathbf{p}^\prime)=2M \left( p_{1\alpha} \delta^\mu_\beta -p_{1\beta} \delta^\mu_\alpha \right)\left( p_{3\sigma}\eta^{ \nu\beta}\delta^\alpha_\rho + p_3^\alpha \delta_\sigma^\beta \delta_\rho^\nu\right)\int \d^3 \mathbf{x}\, \Phi \, l^\sigma l^\rho  e^{i(\mathbf{p}-\mathbf{p}^\prime)\cdot \mathbf{x}}\,,
\end{align}
and the amplitude is simply recovered by contracting $ V_\gamma^{\mu\nu}(\mathbf{p},\mathbf{p}^\prime)$ with the polarization tensors of the external photons 
\begin{align}\label{eq:TphoFT}
    T_\gamma(\mathbf{p},\mathbf{p}^\prime)&=\epsilon_{3\mu}^*V_\gamma^{\mu\nu}(\mathbf{p},\mathbf{p}^\prime)\epsilon_{1\nu}\\&=2M \left( p_{3\alpha} \epsilon_{3\beta}^* -p_{3\beta} \epsilon_{3\alpha}^* \right)\left( p_{1\sigma}\epsilon_1^{\beta}\delta^\alpha_\rho+p_1^\alpha \epsilon_{1\rho}\delta_\sigma^\beta \right)\int \d^3 \mathbf{x}\, \Phi \, l^\sigma l^\rho  e^{i(\mathbf{p}-\mathbf{p}^\prime)\cdot \mathbf{x}}\notag\,,
\end{align}
where $\epsilon_1$ ($\epsilon_3^*$) is the polarization tensor of the spin-1 state with momentum $p_1$ ($p_3$). 

Notice that the structure of the amplitude \eqref{eq:TphoFT} is highly reminiscent of its scalar analogue \eqref{eq:Vscll}. We can therefore apply the same strategy: we use the projectors \eqref{eq:projdef} to rewrite the Fourier transform as in (\ref{eq:projll}). Also in this case we are able to resum the solution into exponential functions, and the amplitude for a photon on a Kerr background in the Born regime at $\mathcal{O}(G)$ is 
\begin{align}\label{eq:TphotonKerr}
T_\gamma(p_1,p_3)&=-  \frac{32\pi GM^2}{q^2}\cosh(a \cdot q)\bigg[\Big( (u\cdot p_1) \, (u\cdot p_3)+\frac{q^2}{4}\Big)\epsilon_1\cdot \epsilon_3^* \notag\\
 +&(u\cdot p_1) \left(  q^\nu u^\mu-q^\mu u^\nu\right)\epsilon_{1\mu}\epsilon_{3\nu}^*-\frac{1}{2}\left(q^\mu q^\nu +q^2 u^\mu u^\nu \right)\epsilon_{1\mu}\epsilon_{3\nu}^*\bigg]\notag\\
 -&\frac{32i\pi GM^2}{q^2}\frac{\sinh{(a\cdot q)}}{(a\cdot q)}\bigg[ \epsilon_{\alpha\beta\mu\nu}u^\alpha a^\beta k^\mu q^\nu \Big( (u\cdot p_1) \,\eta^{\mu\nu} -\frac{1}{2}\left(u^\mu p_1^\nu +u^\nu p_3^\nu\right)\Big)\epsilon_{1\mu} \epsilon_{3\nu}^{*}\notag \\
 +&\frac{1}{2}\epsilon_{\alpha\beta\mu\nu}u^\alpha a^\beta q^\mu \Big(\epsilon_1^\nu \Big( (\epsilon_3^*\cdot p_1)+\frac{q^2}{2}(\epsilon_3^*\cdot u) \Big)+\epsilon_3^{*\nu} \Big( (\epsilon_1\cdot p_3)+\frac{q^2}{2}(\epsilon_1\cdot u) \Big)\Big)\bigg]\notag \\
 -& 16\pi GM^2\frac{(a\cdot q) \cosh{(a\cdot q)}-\sinh{(a\cdot q)}}{(a\cdot q)^3}\bigg[( a\cdot p_1 )\,  p_3^\mu \, a^\nu+(a\cdot p_3) \,  a^\mu p_1^\nu\notag\\
 -&\Big( (a\cdot p_1) (a\cdot p_3)+a^2 \, ( u \cdot p_1 )\, (u\cdot p_3)+ \frac{a^2q^2}{2} \Big)\,\eta^{\mu\nu} \notag \\
+& a^2\, (u\cdot p_1) \left( u^\mu \, p_1^\nu + u^\nu \, p_3^\mu\right)+\frac{q^2}{2}\left(  a^\mu a^\nu+a^2\, u^\mu u^\nu\right)+q^2\, q^\mu q^\nu \bigg]\epsilon_{1\mu}\epsilon_{3\nu}^*\,,
\end{align}
in this case expressed exclusively in terms of 4-vectors, where we recall $p_{2\mu}=p_{4\mu}=Mu_{\mu}$, as expected in the probe limit, and $a^\mu$ and $q^\mu$ are orthogonal to $u^\mu$, $a.u = 0$ and $q.u = 0$. We checked this resummation up to $\mathcal{O}(a^{30})$. 

Notice that this result is crossing symmetric, $p_1 \leftrightarrow - p_3$, where $q = p_3 - p_1$, and gauge invariant (vanishing under the replacements $\epsilon_1\rightarrow p_1$ or $\epsilon_3^*\rightarrow p_3$).  Similarly to the scalar calculation, this would not have been the case without the inclusion of boundary terms at spatial infinity, which here not only restore crossing symmetry but also gauge invariance. 

In the eikonal limit $M\gg \omega\gg |\mathbf{q}|$ of \eqref{eq:TphotonKerr}, we obtain
\begin{equation}
   T_\gamma(p_1,p_3)\xrightarrow[M\gg \omega\gg |\mathbf{q}|]{}= \frac{32 \pi  G M^2}{\mathbf{q}^2} \Big[ \, \omega ^2 \cosh (\mathbf{a} \cdot \mathbf{q})-i \omega \epsilon_{i j l} q_i a_j p_l \frac{\sinh (\mathbf{a} \cdot \mathbf{q})}{(\mathbf{a} \cdot \mathbf{q})}\Big]\, ,
\end{equation}
where $\epsilon_1\cdot \epsilon_3^* \sim 1$, and $\epsilon_1\cdot p_3\sim \epsilon_3\cdot p_1\sim 0$ since the two momenta become collinear. This result matches exactly the eikonal limit of the scalar amplitude (\ref{eq:scEik}), which is of course expected since the null geodesic for a minimally coupled particle is not affected by its spin. 

Similarly to the scalar case, we perfectly match the $t$-pole factorization of \cite{Bautista:2021wfy}, and obtain some additional new contact terms, some of which are necessary to ensure gauge invariance. Notice that the overall structure of contact terms is identical to the scalar answer and guarantees cancellations in the $|\mathbf{q}|\rightarrow 0$ limit. Lastly, the Schwarzschild limit $a\rightarrow 0$ also matches amplitude results, as detailed in Appendix \ref{app:ampphoton}.

\section{Conclusion}\label{sec:conclusion}

In this work we explored the Born regime of $2 \to 2$ scattering amplitudes, which is characterized by the hierarchy $b \sim \lambda \gg \lambda_\text{C}$, where $b$ is the impact parameter, $\lambda$ represents the (de Broglie) wavelength of the scattering particles and $\lambda_\text{C}$ is the Compton wavelength of massive particles in the QFT. This is the regime  relevant for scattering in quantum mechanics but also GW scattering in classical backgrounds. An in-depth discussion of all these scales and regimes was done in Chapter \ref{sec:scales}, including a comparison with the eikonal regime (see Figs. \ref{fig:regimes-table} and \ref{fig:regimes-V}).

We reviewed the EOB framework in QFT pioneered by Todorov \cite{Todorov:1970gr} in Chapter \ref{sec:EOB} which defines an effective potential in terms of scattering amplitudes. With this potential one can establish the EOB Schr{\"o}dinger equation which has been historically used to compute relativistic and radiative corrections to energy levels of bound states \cite{Todorov:1970gr, Rizov:1975tr,Crater:1992xp,Jallouli:1996bu}. Furthermore, in the eikonal regime, the WKB limit of the EOB Schr{\"o}dinger equation leads to the amplitude-action relation \cite{Bern:2021dqo} which allows to compute  relativistic (PM) corrections to classical observables, including bound observables such as the perihelion precession \cite{Kalin:2019rwq,Kalin:2019inp,Cho:2021arx}. An advantage of the EOB Schr{\"o}dinger equation compared to other bound state formalisms in QFT, such as the Bethe-Salpeter equation \cite{PhysRev.84.1232}, is that it has the correct probe limit, reducing to the usual Schr{\"o}dinger or Klein-Gordon equations when one mass is much heavier than the other \cite{Todorov:1970gr}.\footnote{We should also mention the recent work \cite{Adamo:2022ooq} that proposed a classical version of Bethe-Salpeter equation. In this work the authors noted that the exponentiated eikonal amplitude at $O(G^2)$ exhibited some poles in momentum space, which were identified as ``classical" bound states. However, these are outside the regime of validity of the eikonal approximation, as discussed in Chapter \ref{sec:scales}. Besides, there are in principle additional contributions to the potential at $O(G^2)$ which vanish in the eikonal regime, as shown in Chapter \ref{sec:2PM}, and should be relevant for bound state calculations.}

In particular, when one of the scattering particles is massless, the Born regime enforces the probe limit $\omega \ll M$ (coming from $\lambda \gg \lambda_\text{C}$), where $\omega$ is the frequency of the massless particle and $M$ is the mass of the other particle. In this regime, we argue that the EOB Schr{\"o}dinger equation reduces to a classical wave equation. We verify this in section \ref{sec:2PM} by computing the Feynman diagrams for the scattering of two minimally coupled scalars at $\mathcal{O}(G^2)$. We find that, in the Born regime, an additional piece contributes to the gravitational potential (given in   \eqref{eq:2PM_tot}) as compared to the well-known result valid in the eikonal regime, which is relevant for the light bending problem (see e.g. \cite{Bjerrum-Bohr:2014zsa}). We identify this term, emerging from the triangle master integral, as coming from the equations of motion of a scalar wave on a Schwarzschild background, i.e. the Regge-Wheeler equation.

It goes without mention that quantum corrections to the Regge-Wheeler equation (and other classical GW equations) are naturally incorporated in the EOB Schr{\"o}dinger equation \eqref{eq:EOBSchrodinger}, which is defined via the full-fledged quantum scattering amplitude. In the same way that this framework has been used to determine QED corrections to the Hydrogen spectrum, we expect that it can be equally used to compute quantum corrections to the quasinormal mode spectrum of black holes (if this exercise is ever relevant).

In Chapter \ref{sec:Kerr}, we reversed the logic to compute the Compton amplitude of scattering off a Kerr BH, which has arguably evaded a clear computation using QFT methods and subsequent matching with BHPT results \cite{Bautista:2021wfy,Bautista:2022wjf,Bautista:2023sdf}. We wrote down the wave equation in Kerr-Schild coordinates \eqref{eq:KSmetric}, and identified the gravitational potential, see \eqref{eq:Vscll} for the scalar and \eqref{eq:Vphoton} for the photon. We expanded this potential in the BH spin $a$ and performed a Fourier transform, which can be determined exactly in terms of combinatorial factors to virtually any order (see Appendix \ref{app:FT}). We determined the Compton amplitude up to $O(a^{30})$ and our answer resums nicely in terms of exponential functions, see eqs. \eqref{eq:VscX} and \eqref{eq:TphotonKerr} for the scalar and photon cases, respectively. The results match the $t$-channel residue of \cite{Arkani-Hamed:2019ymq,Bautista:2021wfy}, respect all expected properties of crossing symmetry, factorisation and gauge invariance, and reduces to the Schwarzschild solution in the $a\rightarrow 0$ limit (including contact terms).\footnote{A similar calculation yields the scalar $\sqrt\text{Kerr}$ amplitude (see Appendix \ref{app:rootkerr}).} 

Given that the gravitational potential is purely long-range, the presence of contact terms, which are analytic in the momentum transfer $q^\mu$,  deserves further discussion. One source of these contact terms is a result of long-distance pieces in the off-shell potential becoming contact after imposing the on-shell conditions. The second source of contact terms is more unexpected. We find that in deriving the Euler-Lagrange equations for a massless wave on a Kerr background, boundary terms at spatial infinity which are typically dropped give in fact non-zero contributions, see   \eqref{eq:boundexample}.\footnote{ These are not Gibbons-Hawking-York boundary terms \cite{Gibbons:1976ue,York:1972sj}, which are added to the Einstein-Hilbert action in spacetimes with boundaries to ensure a well-defined variational principle. Our boundary terms come after varying the action and are usually dropped in the assumption that the variations vanish at infinity.} These lead to additional contact terms which turn out to be necessary to ensure crossing symmetry and gauge invariance of the Kerr-Compton amplitude. The physical interpretation of those terms is still unclear, and it would be interesting to investigate this further. For instance, is it possible that in the derivation of the Teukolsky equation \cite{1973ApJ...185..635T,1973ApJ...185..649P,1974ApJ...193..443T} these boundary terms also play a role? It is unclear to us whether the BHPT answer, which is in terms of partial waves, leads to a crossing-symmetric and gauge invariant amplitude.

It should be noted, however, that none of the obtained contact terms correspond to dissipative effects. This is because we perturbatively expanded the gravitational potential, and the event-horizon is therefore absent. Our answer can be seen as ``minimal" in the sense that it is independent of the choice of boundary condition at the event-horizon. It would be interesting to see whether our (conservative) Kerr-Compton amplitudes, \eqref{eq:VscX} and \eqref{eq:TphotonKerr}, arise in any way from the classical large spin limit of amplitudes of minimally coupled higher spin massive particles. 

We envision a few extensions of our work. One is of course to repeat the computation of Chapter \ref{sec:Kerr}, which was done for the scalar and photon cases, for the case of the graviton. In fact, the gravitational Kerr-Compton amplitude is the more physically interesting object, which can be ``glued" to itself on the $t$-channel to determine the amplitude for the gravitational scattering of two Kerr BHs (see e.g. the recent work \cite{Chen:2024mmm} and references therein). Scattering of a graviton off a massive body has some differences with respect to the scalar and photon cases. Besides the long-distance contribution coming from the $t$-channel graviton exchange, there are also  $s$-channel and $u$-channel exchanges of the massive body itself which give additional contact pieces (analytic in $t$ terms). It would be interesting to see if these contact terms are also captured by our method. We leave this for future work.

 Another natural future direction is the inclusion of tidal or dissipation effects into our answer. In fact, if we were to continue the exercise done in Chapter \ref{sec:2PM} beyond $O(G^2)$ we would eventually find UV divergences. These would require the addition of non-minimal couplings depending on the nature of the compact objects, be it a black hole, a neutron star or something more exotic \cite{Ivanov:2022qqt}. Therefore, a matching with BHPT results eventually becomes necessary if we want to describe BH scattering. However, even in the Schwarzschild case this can be a challenging problem \cite{Ivanov:2024sds}, mostly due to the lack of a sharp separation between the parts of the BHPT data which correspond to long-distance physics and which pertain to ``near-zone" tidal effects. See e.g. \cite{Ivanov:2022qqt,Ivanov:2024sds,Bautista:2023sdf} for a number of recent interesting proposals. We believe that our understanding in terms of the Born series may also aid in better clarifying this near-far factorization.

Finally, it would be interesting to explore the non-linear corrections to the Teukolsky equation (or other classical GW wave equations). As confirmed in recent numerical GR simulations \cite{Mitman:2022qdl,Cheung:2022rbm}, non-linear quasinormal overtones contribute to the GW waveform in the early ringdown phase. These effects would in principle be captured by higher-point gravitational amplitudes in the Born regime. We expect the Feynman diagram expansion of higher-point amplitudes to also take the form of a (multi-particle) Born series (see e.g. \cite{Jones:2022aji} for work in this context).

\section*{Acknowledgements}
We are grateful to Fabian Bautista, Brando Bellazzini, Zvi Bern, Lucile Cangemi, Vitor Cardoso, Simon Caron-Huot, Stefano De Angelis, Juan Pablo Gatica, Alessandro Georgoudis, Riccardo Gonzo, Callum Jones, Florian Nortier, Julio Parra-Martinez, and Mikhail Solon for useful discussions. We thank Fabian Bautista, Brando Bellazzini, Simon Caron-Huot and Florian Nortier for comments on the draft. M.C.'s work is supported by the Simons Collaboration on the Non-perturbative Bootstrap. G.I. is supported by the US Department of Energy under award number DE-SC0024224, the Sloan Foundation and the Mani L. Bhaumik Institute for Theoretical Physics.

\appendix

\section{Schr{\"o}dinger to Lippmann-Schwinger}
\label{app:LS}
In this Appendix, we derive \eqref{eq:VEOB} and \eqref{eq:TEOB}. In particular, we want to fix their normalisation in order to reproduce the Lippmann-Schwinger equation as defined in \eqref{eq:LS}.

We first Fourier transform the Green's function \eqref{eq:G} 
\begin{align}
  & \int \frac{\d^3 \mathbf{k}}{(2\pi)^3}e^{i\mathbf{k}\cdot \mathbf{x}}G(\mathbf{p},\mathbf{k})\notag\\&  =\frac{1}{16\pi^3\sqrt{s}}\int \frac{\mathbf{k}^2 \d |\mathbf{k}|}{(2\pi)^2}\frac{1}{|\mathbf{k}|^2-|\mathbf{p}|^2-i\epsilon}\int_{-1}^{1} d\mu \, e^{i|i\mathbf{k}||\mathbf{x}|\mu}\notag\\
  &=\frac{1}{16\pi^3 i\sqrt{s}|\mathbf{x}|}\int_0^\infty \frac{|\mathbf{k}| \d |\mathbf{k}|}{(2\pi)^2}\frac{\left( e^{i|\mathbf{k}||\mathbf{x}|}-e^{-i|\mathbf{k}||\mathbf{x}|}\right)}{|\mathbf{k}|^2-|\mathbf{p}|^2-i\epsilon}\notag\\
  &=\frac{1}{4\sqrt{s}}\frac{e^{i|\mathbf{p}||\mathbf{x}|}}{(2\pi )^4|\mathbf{x}|}\,,
  \end{align}
which implies \eqref{eq:GX}.

The Green function in position space $G(\mathbf{x},\mathbf{x}^\prime)$ satisfies 
\begin{align}
(\nabla^2+|\mathbf{p}|^2)G(\mathbf{x},\mathbf{x}^\prime)=\delta^3(\mathbf{x}-\mathbf{x}^\prime)\, ,
\end{align}
and therefore it can be employed to solve the EOB Schr{\"o}dinger equation \eqref{eq:EOBSchrodinger}, as 
\begin{align}\label{eq:EOBSolG}
    \Psi_{\mathbf{p}}(\mathbf{x})&=\Psi_{\mathbf{p}}^{(0)}(\mathbf{x})+\int \d^3 \mathbf{x}^\prime \, G(\mathbf{x},\mathbf{x}^\prime) \,V( \mathbf{x}^\prime) \, \Psi_{\mathbf{p}}(\mathbf{x}^\prime) \notag \\
    &=e^{i\mathbf{p}\cdot \mathbf{x}}-\frac{1}{4\pi}\int \d^3\mathbf{x}^\prime \frac{e^{i|\mathbf{p}||\mathbf{x}-\mathbf{x}^\prime|}}{|\mathbf{x}-\mathbf{x}^\prime|}V(\mathbf{x}^\prime) \, \Psi_{\mathbf{p}}(\mathbf{x}^\prime)\,,
\end{align}
where we consider the incoming wave to be a plane wave $\Psi_{\mathbf{p}}^{(0)}(\mathbf{x})=e^{i\mathbf{p}\cdot \mathbf{x}}$. So far the solution is exact.

Now we consider that the potential $V(\mathbf{x})$ obeys a power law decay with distance, so that the integral in \eqref{eq:EOBSolG} has approximately a compact support. We then take the large $|\mathbf{x}|$ limit
\begin{equation}
    |\mathbf{x}-\mathbf{x}^\prime|\approx |\mathbf{x}|-\frac{\mathbf{x}\cdot \mathbf{x}^\prime}{|\mathbf{x}|}=|\mathbf{x}|-\hat{\mathbf{x}} \cdot \mathbf{x}^\prime\, ,
\end{equation}
 and define $\mathbf{p}^\prime=|\mathbf{p}| \hat{\mathbf{x}}$. 
 In this limit, the solution \eqref{eq:EOBSolG} becomes
\begin{align}\label{eq:EOBSolG2}
    \Psi_{\mathbf{p}}(\mathbf{x} \to \infty)&=e^{i\mathbf{p}\cdot \mathbf{x}}-\frac{1}{4\pi}\frac{e^{i|\mathbf{p}||\mathbf{x}|}}{|\mathbf{x}|}\int \d^3\mathbf{x}^\prime e^{-i\mathbf{p}^\prime\cdot\mathbf{x}^\prime}V(\mathbf{x}^\prime) \, \Psi_{\mathbf{p}}(\mathbf{x}^\prime)\,.
\end{align}
From this expression we can extract the scattering amplitude, as the coefficient of the outgoing wave
\begin{equation}\label{eq:TEOB2}
    T(\mathbf{p}, \mathbf{p}') = N \int \d^3 \mathbf{x}^\prime \, e^{-i \mathbf{p}' \cdot \mathbf{x}^\prime } \, V(\mathbf{x}^\prime) \,\Psi_{\mathbf{p}}(\mathbf{x}')\,,
\end{equation}
where $N$ is a normalisation factor that we are going to fix immediately by matching to the Lippmann-Schwinger equation \eqref{eq:LS}.

We apply $N\int \d^3 \mathbf{x}\, e^{-i\mathbf{p}^\prime\cdot \mathbf{x}}V(\mathbf{x})$ on both sides of \eqref{eq:EOBSolG} and by using the amplitude definition \eqref{eq:TEOB}, we recover
\begin{align}\label{eq:LSalmost}
     T(\mathbf{p}, &\mathbf{p}^\prime)=N\int \d^3\mathbf{x}\, e^{-i\mathbf{p}^\prime\cdot \mathbf{x}}V(\mathbf{x})e^{i\mathbf{p}\cdot \mathbf{x}}\\&+\int \d^3\mathbf{k}\,G(\mathbf{p},\mathbf{k})\,\left[-2\sqrt{s}\int\d^3\mathbf{x}\, e^{-i\mathbf{p}^\prime\cdot \mathbf{x}}V(\mathbf{x})e^{i\mathbf{k}\cdot \mathbf{x}}\right]\left[N\int \d^3\mathbf{x}^\prime e^{-i\mathbf{k}\cdot \mathbf{x}^\prime}\Psi_\mathbf{p}(\mathbf{x}^\prime)\right]\notag\\
     &=N\int \d^3\mathbf{x}\, e^{-i\mathbf{p}^\prime\cdot \mathbf{x}}V(\mathbf{x})e^{i\mathbf{p}\cdot \mathbf{x}}\notag\\&+\int \d^3\mathbf{k}\,G(\mathbf{p},\mathbf{k})\,\left[-2\sqrt{s}\int\d^3\mathbf{x}\, e^{-i\mathbf{p}^\prime\cdot \mathbf{x}}V(\mathbf{x})e^{i\mathbf{k}\cdot \mathbf{x}}\right]T(\mathbf{p},\mathbf{k})\notag\,,
\end{align}
 where we replaced the Green's functions $G(\mathbf{x},\mathbf{x}^\prime)$ by its conjugate \eqref{eq:GX}. In order to match the Lippmann-Schwinger equation, we define
 \begin{align}
     V(\mathbf{p},\mathbf{p}^\prime)=-2\sqrt{s}\int \d^3\mathbf{x}\, e^{-i\mathbf{p}^\prime\cdot \mathbf{x}}V(\mathbf{x})e^{i\mathbf{p}\cdot \mathbf{x}}\,,
 \end{align}
 and fix the normalisation $N$ to 
 \begin{equation}
     N=-2\sqrt{s}\,.
 \end{equation}
Applying these definitions to \eqref{eq:LSalmost}, we obtain the Lippmann-Schwinger equation
\begin{equation}
    T(\mathbf{p},\mathbf{p}^\prime)=V(\mathbf{p},\mathbf{p}^\prime)+\int \d^3\mathbf{k}\,T(\mathbf{p},\mathbf{k})G(\mathbf{p},\mathbf{k})V(\mathbf{k},\mathbf{p}^\prime)\, ,
\end{equation}
given in \eqref{eq:LS}.

\section{Iteration pieces in the Born series}\label{app:iteration}
In this Appendix, we show that in the Born regime the combination of box and cross-box master integrals in \ref{eq:oneloopIBP} combines to generate the iteration of the leading order potential. We restrict to the $2\rightarrow 2$ kinematic of \ref{eq:kinp1p2} and \ref{eq:kinp3p4} with one massless and one massive state. In this derivation we follow closely \cite{Cheung:2018wkq}, despite the fact that we have an external massless particle and the relevant limit is $M\gg \omega\sim |\mathbf{q}|$, as opposed to $m_i\sim p_i\gg |\mathbf{q}|$.

The two master integrals are given by
\begin{align}\label{eq:BoxXbox}
     \mathcal{I}_\Box &=\int \frac{\d^4l}{(2\pi)^4} \frac{1}{(l-p_1)^2} \frac{1}{(l-p_3)^2} \frac{1}{l^2} \frac{1}{(p_1+p_2-l)^2-M^2}\,
     , \notag\\
      \mathcal{I}_{\mathrm{X}}&=\int \frac{\d^4l}{(2\pi)^4} \frac{1}{(l-p_1)^2} \frac{1}{(l-p_3)^2} \frac{1}{l^2} \frac{1}{(p_2-p_3+l)^2-M^2}
      ,.
\end{align}
We find convenient to work with the following variables
\begin{equation} \label{eq:defXY}
    X_1=|p-l_1|,\quad X_2=|k-l_1|,\quad Y_1=l_1^2-\omega^2
\end{equation}
where $l=(\omega+l_0,l_1)$. 

Now we explicitly evaluate the $l_0$ component of the loop momenta in (\ref{eq:BoxXbox}) by deforming the contour both in the lower and upper plane, effectively averaging over all residues
\begin{equation}\label{eq:contour}
    \int \frac{\d l_0}{2\pi}(.)=\frac{i}{2}\left[\sum_{l_0^*\in H^+} \text{Res}(.)-\sum_{l_0^*\in H^-} \text{Res}(.) \right]
    , 
\end{equation}
where $H^\pm$ are respectively the upper and lower half planes.

We first explicitly separate the time component obtaining for instance for the box
\begin{align}
   \mathcal{I}_\Box=\int \frac{\d^3l_1}{(2\pi)^3}\int \frac{\d l_0}{2\pi}& \frac{1}{l_0^2-X_1^2+i 0} \frac{1}{l_0^2-X_2^2+i 0}\notag\\& \frac{1}{l_0^2+2\omega l_0-Y_1+i 0} \frac{1}{l_0^2-2\sqrt{M^2+\omega^2} l_0-Y_1+i 0} \, .
\end{align}
Notice that from (\ref{eq:coeffIBP}) the overall triangle scaling $c_\triangle \mathcal{I}_\triangle\sim M^4/M\sim M^3$ and since $c_\Box=c_\mathrm{X}\sim M^4$, we focus only on the $1/M$ contribution of $\mathcal{I}_\Box$ and $\mathcal{I}_\mathrm{X}$.

By direct inspection of the 8 residues picked up by the integration contour (\ref{eq:contour}), we observe that despite the fact that all poles of graviton propagators scale as $1/M$ they perfectly cancel between box and cross-box. Same destiny goes for all other residues, that cancel pairwise between box and cross-box, except for the matter pole resulting from the heavy mass propagator. In this case, the two equal contributions sum up leading to
\begin{equation}
    \mathcal{I}_\Box + \mathcal{I}_\mathrm{X}=\frac{i}{2M}\int \frac{\d^3 l_i}{(2\pi)^3}\frac{1}{X_1^2 Y X_2^2}+\mathcal{O}\left(\frac{1}{M^2}\right)\,.
\end{equation}
Replacing the variables defined in (\ref{eq:defXY}), we recover the result of (\ref{eq:2PM_tot}).

\section{Choice of coordinate system and off-shell pieces}\label{app:schw_coord}

In this section, we illustrate how a different coordinate choice which appears vastly different off-shell leads to the same on-shell potential, by studying the explicit example of Schwarzschild coordinates (as opposed to isotropic coordinates, employed in the main text).

The metric we consider is
\begin{equation}
    ds^2 = f(r) dt^2 - {1 \over f(r)} dr^2 - r^2 d \theta^2 - r^2 \sin^2 \theta \,d \varphi^2    
\end{equation}
with $f(r) \equiv 1 - 2 G M/r$. 

We extract the potential from the wave equation, repeating the steps of section (\ref{sec:RW}) obtaining
\begin{equation}
V(r) = - \left(1 - {1 \over f(r)} \right) \hat{E}^2 + \big(f(r) - 1 \big) \partial_r^2 + \bigg( f'(r) + 2 {f(r) - 1 \over r} \bigg) \partial_r
\end{equation}
which at $O(G)$ gives
\begin{align}
\label{eq:VSchwar}
V(r) &=  {2 G M \over r} \hat{E}^2 - {2 G M \over r^2} \partial_r - {2 G M \over r} \partial_r^2 \notag \\
&=  {2 G M \over r} \hat{E}^2 - {2 G M \over r^3} (\mathbf{r} \cdot \partial_\mathbf{r}) - {2 G M \over r} \left({\mathbf{r} \cdot \partial_\mathbf{r} \over r}\right)^2 +\mathcal{O}(G^2)
\end{align}
where 
\begin{align}
    \left({\mathbf{r} \cdot \partial_\mathbf{r} \over r}\right)^2 \phi(r) = \left({ r_i  \partial_i \over r}\right) \left({r_j \partial_j \over r}\right) \phi(r) = \left({ r_i r_j  \over r^2}\right)  \partial_i \partial_j \phi(r) 
\end{align}
We now want to perform the Fourier transform of
\begin{equation}
    \int d^{d} \mathbf{r} \; e^{- i \mathbf{p} \cdot \mathbf{r} } \,V(r)\, \phi(r).
\end{equation}
We write
\begin{equation}
    V(r) =  \int {\d^{d} \mathbf{q} \over (2 \pi)^d} e^{ i \mathbf{k} \cdot \mathbf{r} } \,V(\mathbf{q}), \qquad \phi(\mathbf{r}) =  \int {\d^{d} \mathbf{q} \over (2 \pi)^d} e^{i \mathbf{k} \cdot \mathbf{r} } \phi(\mathbf{q}).
\end{equation}
The first piece of the potential in \eqref{eq:VSchwar} gives
\begin{align}
     &\int \d^{3} \mathbf{r} \; e^{- i \mathbf{p} \cdot \mathbf{r} } \,V(r)\, \phi(r) = {2 G M \omega^2}    \int {\d^{3} \mathbf{k} \over (2 \pi)^3} \phi(\mathbf{k}) \int \d^{3} \mathbf{r} \; e^{i(\mathbf{k} - \mathbf{p}) \cdot \mathbf{r} } {1 \over r} \,\notag \\
     &= {2 G M \omega^2 \over 2 \pi^2} \int \d^3 \, \mathbf{k} \, \phi(\mathbf{k}) \, {1 \over |\mathbf{k} - \mathbf{p}|^2}
\end{align}
where we made use of the identity
\begin{equation}
    \int {\d^{3} \mathbf{r} \over (2 \pi)^3} e^{- i \mathbf{q} \cdot \mathbf{r} } {1 \over r^{2 \nu}} = \frac{2^{-2 \nu } \Gamma \left(\frac{3}{2}-\nu \right) }{\pi ^{3/2} \Gamma (\nu )} {1 \over |\mathbf{q}|^{3 - 2 \nu} }
\end{equation}
The second piece in \eqref{eq:VSchwar} gives
\begin{align}
     &- 2 G M \int \d^{3} \mathbf{r} \; e^{- i \mathbf{p} \cdot \mathbf{r} } \, {\mathbf{r} \cdot \partial_\mathbf{r} \phi(r) \over r^3} = -
     {2 G M }    \int {\d^{3} \mathbf{k} \over (2 \pi)^3} \phi(\mathbf{k}) \, (\mathbf{k} \cdot \partial_\mathbf{k}) \int \d^{3} \mathbf{r} \;   { e^{i(\mathbf{k} - \mathbf{p}) \cdot \mathbf{r} } \over r^3} \, \notag \\
     &= {2 G M \over 2 \pi^2} \int \d^3 \, \mathbf{k} \, \phi(\mathbf{k}) \, (\mathbf{k} \cdot \partial_\mathbf{k} \log |\mathbf{k} - \mathbf{p}| ) = {2 G M \over 2 \pi^2} \int \d^3 \, \mathbf{k} \, \phi(\mathbf{k}) \, {\mathbf{k} \cdot (\mathbf{k} - \mathbf{p} ) \over |\mathbf{k} - \mathbf{p}|^2 }
\end{align}
The third piece in \eqref{eq:VSchwar} gives 
\begin{align}
     &- 2 G M \int \d^{3} \mathbf{r} \; e^{- i \mathbf{p} \cdot \mathbf{r} } \, {r_i r_j  \partial_i \partial_j \phi(\mathbf{r}) \over r^3} = -
     {2 G M }  \!  \int {\d^{3} \mathbf{k} \over (2 \pi)^3} \phi(\mathbf{k}) \, k_i k_j \partial_{k_i} \partial_{k_j} \! \int \d^{3} \! \mathbf{r} \;   { e^{i(\mathbf{k} - \mathbf{p}) \cdot \mathbf{r} } \over r^3} \, \notag \\
     &= {2 G M \over 2 \pi^2} \int \d^3 \, \mathbf{k} \, \phi(\mathbf{k}) \, k_i k_j \partial_{k_i} \partial_{k_j} \!\log |\mathbf{k} - \mathbf{p}|  = {2 G M \over 2 \pi^2} \int \d^3 \, \mathbf{k} \, \phi(\mathbf{k}) \, k_i k_j \partial_{k_i} { (k_j -p_j)\over |\mathbf{k} - \mathbf{p}|^2} \notag  \\
    &= {2 G M \over 2 \pi^2} \int \d^3 \, \mathbf{k} \, \phi(\mathbf{k}) \, k_i k_j  {  \delta_{ij} - 2 (k_i - p_i)(k_j - p_j)/|\mathbf{k} - \mathbf{p}|^2 \over |\mathbf{k} - \mathbf{p}|^2} \notag \\
    &= {2 G M \over 2 \pi^2} \int \d^3 \, \mathbf{k} \, \phi(\mathbf{k}) \,  {  |\mathbf{k}|^2 |\mathbf{k} - \mathbf{p}|^2 - 2 \big[\mathbf{k} \cdot (\mathbf{k} - \mathbf{p}) \big]^2 \over |\mathbf{k} - \mathbf{p}|^4} \notag \\
\end{align}
We can now read off the potential via
\begin{equation}
    \int \d^{3} \mathbf{r} \; e^{- i \mathbf{p} \cdot \mathbf{r} } \,V(r)\, \phi(\mathbf{r}) = \int {\d^{3} \mathbf{k} \over (2 \pi)^3} \,V(\mathbf{p}, \mathbf{k})\, \phi(\mathbf{k}) 
\end{equation}
and we have at $O(G)$
\begin{equation}
    V(\mathbf{p}, \mathbf{k}) = 8 \pi G M \left[ {\omega^2 \over |\mathbf{q}|^2} + {\mathbf{p} \cdot \mathbf{q} + |\mathbf{q}|^2 \over |\mathbf{q}|^2} + {|\mathbf{p}|^2 \over |\mathbf{q}|^2} -  {2(\mathbf{p} \cdot \mathbf{q} + |\mathbf{q}|^2)^2 \over |\mathbf{q}|^4} \right]
\end{equation}
where we set $\mathbf{k} = \mathbf{p} + \mathbf{q}$.
We can see that on the mass-shell, where $|\mathbf{k}| = |\mathbf{p}|$, i.e. $2 \mathbf{p} \cdot \mathbf{q} = - |\mathbf{q}|^2$ we recover
\begin{equation}
    V(\omega, |\mathbf{q}|) = {16 \pi G M \omega^2 \over |\mathbf{q}|^2} + O(G^2)
\end{equation}
which reproduces the $\mathcal{O}(G)$ piece of (\ref{eq:V01pm}).

\section{Photon gravitational potential from Feynman diagrams}\label{app:ampphoton}
In this Appendix we extract the gravitational photon potential at $\mathcal{O}(G)$ from a standard tree-level computation. In this case we only need to consider one diagram with one graviton exchange, therefore the amplitude can be simply constructed by gluing two 3-point functions. We use the kinematic of \eqref{eq:kinp1p2} and \eqref{eq:kinp3p4}.

The minimal coupling between two incoming photons with momenta $k$ and $k^\prime$ and a graviton is given by
 \begin{equation}\label{eq:photgrav}
 (8\pi G)^{1/2}\left[\left(k_{[\mu} \epsilon_{\alpha]}k^{\prime}_{[\nu} \epsilon^{\prime}_{\beta]}+ k_{[\nu} \epsilon_{\alpha]}k^{\prime}_{[\mu} \epsilon^{\prime}_{\beta]} \right)\eta^{\alpha\beta}
 -\frac{1}{2}\eta_{\mu\nu}k ^{[\alpha}\epsilon ^{\beta]} k^\prime_{[\alpha}\epsilon^{\prime}_{\beta]}\right]\,,
 \end{equation}
 while the minimal coupling of two incoming massive scalars of mass $M$ with momenta $k$ and $k^\prime$ to gravity is
 \begin{equation}\label{eq:scagrav}
     (8\pi G)^{1/2}\left(k^\mu k^{\prime\nu}+k^{\prime\nu} k^\mu- \eta^{\mu\nu}((k\cdot k^\prime)+M^2)\right)\,.
 \end{equation}
 Finally, the graviton propagator is 
 \begin{equation}\label{eq:propgrav}
     \frac{1}{2}\left(\eta^{\mu\alpha}\eta^{\nu\beta}+\eta^{\mu\beta}\eta^{\nu\alpha}-\frac{1}{2}\eta^{\mu\nu}\eta^{\alpha\beta}\right)\,.
 \end{equation}
 The amplitude is simply given by the product of \eqref{eq:photgrav}, \eqref{eq:scagrav} and \eqref{eq:propgrav} evaluated on the kinematics \eqref{eq:kinp1p2} and \eqref{eq:kinp3p4}
\begin{align}
 T(s,t)=- \frac{2}{M_{pl}^2}\epsilon_1^\mu \epsilon_3^{*\nu} &\Big[ \frac{(M^4-s u)}{t}g_{\mu\nu}+\frac{(M^2-s)}{t}p_{2\mu}p_{1\nu}+\frac{(M^2-u)}{t}p_{4\mu}p_{1\nu}\notag\\
  &+\frac{(M^2-u)}{t}p_{3\mu}p_{2\nu}+\frac{(M^2-s)}{t}p_{3\mu}p_{4\nu}+\left(1-\frac{2M^2}{t}\right)p_{3\mu}p_{1\nu}\notag\\
 &+ p_{2\mu}p_{4\nu} +p_{4\mu}p_{2\nu}\Big]  
\end{align}
We rescale now $p_2=M u_2$ and $p_4=M u_2$, where in the large mass limit $u_{2\mu}=(1,0,0,0)$. The expansion at large $M$ becomes
\begin{equation}
 T(p_1,p_3)=- 16 \pi G M^2\epsilon_1^\mu \epsilon_3^{*\nu} \Big[ \left(1+\frac{4\omega^2}{t}\right)\eta_{\mu\nu}+\frac{4\omega}{t}\left(q_{\mu} u_{2\nu}-u_{2\mu} q_{\nu}\right)
-\frac{2q_{\mu}q_{\nu}}{t}-2u_{2\mu}u_{2\nu}\Big]  \,,
\end{equation}
which matches \ref{eq:TphotonKerr}, by taking the $a\rightarrow 0$ limit.

\section{$\sqrt{\mathrm{Kerr}}$ amplitude}\label{app:rootkerr}
In this appendix, we show that the same technology used to extract Kerr amplitudes in section \ref{sec:Kerr}, can straightforwardly be applied to the case of scalar root Kerr, where the wave equation takes the form
\begin{equation}
    D_\mu D^\mu \phi =0\,, \qquad D_\mu = \partial_\mu + i e \,\Phi_{\sqrt{\mathrm{Kerr}}}\, l_\mu\,, \quad \Phi_{\sqrt{\mathrm{Kerr}}}=\frac{r^3}{r^4+(a\cdot x)^2}
\end{equation}
By repeating the argument of Kerr (including boundary terms), we simply obtain 
\begin{align}
    V_{\sqrt{\mathrm{Kerr}}}&=i e (p_\mu + k_\mu ) \int d^3 x \, \Phi l^\mu e^{i(k-p)x}\notag\\
   & =ie (p_\mu + k_\mu ) \left[ u^\mu \int d^3 x \, \Phi e^{i(k-p)x} + \Pi_T^{\mu i}\int d^3  x \, \Phi l_i e^{i(k-p)x}\, \right]\notag\\
   &=\frac{8\pi ie }{q^2}\left[\omega\cosh{aq}-i \epsilon_{\mu \nu\alpha\beta}a^\mu k^\nu q^\alpha u^\beta \frac{\sinh{aq}}{aq}\right]\, .
\end{align}
In this case, the amplitude does not display any contact term, and therefore it matches the eikonal computation of e.g. \cite{Alessio:2023kgf}.

\section{Fourier transforms}\label{app:FT}

The general Fourier transform we need to do is of the form
\begin{equation}
    {{x_{i_1} \cdots x_{i_n}} \over r^m} \partial_{j_1} \cdots \partial_{j_{n'}}
\end{equation}
where $r^2 \equiv x^i x^i$ and where most of its indices will be contracted with $a^i$ the black hole spin.
\par
We have
\begin{equation}
    \int d^3 x \, e^{ - i \vec{p} \cdot \vec{x}} \, {{x_{i_1} \cdots x_{i_n}} \over r^m} \partial_{j_1} \cdots \partial_{j_{n'}} \phi(\vec{x})
\end{equation}
In terms of the Fourier transform of the field $\phi(k)$ we have
\begin{equation}
    \phi(\vec{x}) = {1 \over (2 \pi)^3} \int d^3k \, e^{i \vec{k} \cdot \vec{x}} \phi(\vec{k})
\end{equation}
so that the Fourier transform reads (where $\vec{q} \equiv \vec{k} - \vec{p}\;$)
\begin{align}
   &{1 \over (2 \pi)^3} \int d^3k \, \phi(\vec{k}) \; (i k^{j_1} )\cdots (i k^{j_n'} ) \int d^3 x \, e^{i \vec{q} \cdot \vec{x}} \, {{x_{i_1} \cdots x_{i_n}} \over r^m}\notag \\
   &= {1 \over (2 \pi)^3} \int d^3k \, \phi(\vec{k}) \; (i k^{j_1} )\cdots (i k^{j_n'} ) (-i \partial_{q_{i_1}}) \cdots (-i \partial_{q_{i_n}}) \int d^3 x \, {e^{i \vec{q} \cdot \vec{x}} \over r^m} 
\end{align}
The potential is read off from 
\begin{equation}
 \int d^{3} \mathbf{r} \; e^{- i \mathbf{p} \cdot \mathbf{r} } \,V(r)\, \phi(\mathbf{r}) = \int {d^{3} \mathbf{k} \over (2 \pi)^3} \,V(\mathbf{p}, \mathbf{k})\, \phi(\mathbf{k}) 
 \end{equation}
 so that a contribution of the form
\begin{equation}
   V(\vec{x}) = {{x_{i_1} \cdots x_{i_n}} \over r^m} \partial_{j_1} \cdots \partial_{j_{n'}}
\end{equation}
will contribute to $V(\vec{p},\vec{k})$ as
\begin{equation}
V(\vec{p},\vec{k}) = (-i)^{n} (i)^{n'} \; k^{j_1}  \cdots  k^{j_{n'}}  \; \partial_{q_{i_1}} \cdots \partial_{q_{i_n}} \int d^3 x \, {e^{i \vec{q} \cdot \vec{x}} \over r^m} 
\end{equation}

\subsection{Scalar isotropic Fourier transforms}

The most basic Fourier transform we need is the following
\begin{equation}
   V_m(q)= \int {d^d x} \, {e^{i \vec{q} \cdot \vec{x}} \over r^m} = \pi^{\frac{3}{2}-\epsilon }  
 \frac{ \Gamma \left(\frac{3-m}{2}-\epsilon \right)}{\Gamma \left(\frac{m}{2}\right)} \bigg({2 \over |\vec{q}\,|}\bigg)^{3 - m - 2 \epsilon}, \; \; d = 3 - 2 \epsilon
\end{equation}
For example, for $m = 1$ we have
\begin{equation}
    V_{m}(q) = {4 \pi \over q^2}, \qquad m = 1.
\end{equation}
For $m = 2,4,6,8,...$ even:
\begin{align}
    V_m(q) = \frac{2 \pi ^2 (-1)^{m-2 \over 2}}{\Gamma (m-1)} q^{m-3}, \qquad m = 2,4,6,8,10,...
\end{align}
For $m =3,5,7,9$ odd we need to expand in $1/\epsilon$ and find a formally divergent expression
\begin{equation}
V_m(q) = {2 \pi  (-1)^{\frac{m+1}{2}} \over \Gamma (m-1)}  \left(\psi ^{(0)}_\frac{m-1}{2}-\frac{1}{\epsilon }+\log (4 \pi ) -2 \log (q)\right) q^{m-3}
\label{eq:Vmq}
\end{equation}
where $\psi^{(0)}$ is the digamma function.
\par
Let us now compute the $n$-th derivative $V_m^{(n)}(q)$. 
\par
\noindent For $m = 1$ we have
\begin{align}
V_1^{(n)} = 4 \pi  (-2)^{(n)} q^{-n-2}, \qquad m = 1
\end{align}
where $x^{(n)} \equiv \frac{\Gamma (x+1)}{\Gamma (-n+x+1)}$ is the factorial power.
\par
\noindent For even $m = 2,4,6,8,...$ 
\begin{align}
    V_m^{(n)}(q) = \frac{2 \pi ^2 (m-3)^{(n)}  (-1)^{m-2 \over 2}}{\Gamma (m-1)} q^{m-n-3}, \qquad m = 2,4,6,8,10,...
\end{align}
For odd $m =3,5,7,9,...$ we have
\begin{align}
    V_m^{(n)}(q) &= {2 \pi  (-1)^{\frac{m+1}{2}} \over \Gamma (m-1)}   q^{m-n-3} \notag \\
    &\times \bigg[\left(\psi ^{(0)}_\frac{m-1}{2}-\frac{1}{\epsilon }+\log (4 \pi ) -2 \log (q)\right) (m-3)^{(n)}
     + 2 (-1)^{n} c_{m-3,n} \bigg], \notag \\
     &\qquad \qquad \qquad \qquad \qquad \qquad \qquad \qquad \qquad \qquad m = 3,5,7,9,...
\end{align}
with
\begin{align}
    c_{m,n} &\equiv  \,(-m)_{m} \, (1)_{n-m-1} \;, \qquad \text{ if } n > m \\
    &\equiv \, -(-m)_n \,H_{m,n} \;, \qquad  \;\;\; \; \, \text{ if } n \leq m
\end{align}
where $(x)_n$ is the Pochhammer symbol and 
\begin{equation}
H_{m,n} \equiv \sum_{j=0}^{n-1} {1 \over m - j}
\end{equation} 
we call the \emph{Harmonic difference}. 

\subsection{Tensorial isotropic Fourier transforms}
Now we are interested in taking derivatives $\partial_{q_{i_1}} \cdots \partial_{q_{i_n}}$ of the scalar Fourier transforms. If $f(q) = f(|\vec{q}|)$ is a function only of the modulus, we can re-express the gradient $n$-th derivative $\partial_{q_{i_1}} \cdots \partial_{q_{i_n}} f(q)$ in terms of the tensor structures built out of $q_i$ and $\delta_{i j}$ given that $f(q)$ only depends on the norm. We find
\par
\noindent If $n$ is even:
\begin{align}
    \partial_{q_{i_1}} \!\cdots \partial_{q_{i_n}} f(q) &= q_{i_1} \!\cdots q_{i_n} \,f_{(n)}(q) \notag \\
    & + \delta_{(i_1 i_2} q_{i_3} \!\cdots q_{i_n)} \,f_{(n-1)}(q) \notag \\
    & + \delta_{(i_1 i_2} \delta_{i_3 i_4} q_{i_5} \!\cdots q_{i_n)} \,f_{(n-2)}(q) \notag \\
    & + \dots \notag \\
     & + \delta_{(i_1 i_2} \! \cdots \delta_{i_{n-3} i_{n-2}} \, q_{i_{n-1}} q_{i_n)} f_{({n + 2\over 2})}(q) \notag \\
    & + \delta_{(i_1 i_2} \! \cdots \delta_{i_{n-1} i_n)} \, f_{({n \over 2})}(q)
\end{align}
If $n$ is odd:
\begin{align}
    \partial_{q_{i_1}} \!\cdots \partial_{q_{i_n}} f(q) &= q_{i_1} \!\cdots q_{i_n} \,f_{(n)}(q) \notag \\
    & + \delta_{(i_1 i_2} q_{i_3} \!\cdots q_{i_n)} \,f_{(n-1)}(q) \notag \\
    & + \delta_{(i_1 i_2} \delta_{i_3 i_4} q_{i_5} \!\cdots q_{i_n)} \,f_{(n-2)}(q) \notag \\
    & + \dots \notag \\
     & + \delta_{(i_1 i_2} \! \cdots \delta_{i_{n-4} i_{n-3}} \, q_{i_{n-2}} q_{i_{n-1}} q_{i_n)} f_{({n + 3 \over 2})}(q) \notag \\
    & + \delta_{(i_1 i_2} \! \cdots \delta_{i_{n-2} i_{n-1}} \, q_{i_n)} \, f_{({n + 1 \over 2})}(q)
\end{align}
or, in compact form,
\begin{equation}
  \partial_{q_{i_1}} \!\cdots \partial_{q_{i_n}} f(q) = \sum_{j=0}^{\lfloor {n \over 2} \rfloor}  \delta_{(i_1 i_2} \cdots \delta_{i_{2j-1} i_{2j}} \, q_{i_{2j + 1}} \cdots \, q_{i_n)} \, f_{(n-j)}(q)
\end{equation}
where ${\lfloor {n \over 2} \rfloor} = {n \over 2}$ if $n$ is even and ${\lfloor {n \over 2} \rfloor} = {n - 1 \over 2}$ if $n$ is odd.
\par
We define symmetrization over a string of $\delta$'s and $q$'s as follows
\begin{equation}
    \delta_{(i j} \! \cdots \delta_{k l} \, q_{m} \cdots q_{n)} \equiv {1 \over 2^{n_\delta} n_\delta!  n_q !} \sum_\text{permutations} \delta_{i j} \! \cdots \delta_{k l} \, q_{m} \cdots q_{n}
\end{equation}
where $n_\delta$ and $n_q$ are the number of $\delta$'s and $q$'s respectively.\footnote{The reason for this definition is that ${n! \over 2^{n_\delta} n_\delta!  n_q !}$ is the number of permutation invariant terms (the $2^{n_\delta}$ term accounts for the symmetry of the Kronecker delta $\delta_{i j} = \delta_{j i}$). In this way, after symmetrization there will be no overall numerical coefficient produced in front of the tensors.}
\par
The derivative coefficients $f_{(n)}$ obey the recursion relation 
\begin{equation}
    f_{(n+1)} = {1 \over q} {d f_{(n)} \over d q}
\end{equation}
and are given by
\begin{equation}
    f_{(n)}(q) \equiv  \, \sum_{j = 1}^n   \, {b_{n,j}\over q^{n + j -1}} \, {d^{n-j + 1} f(q) \over d q^{n-j + 1}} 
\end{equation}
with
\begin{align}
  b_{n,j} =  2^{1-j} \frac{ \, (1-n)_{j-1} \, (n)_{j-1}}{\Gamma(j)} =
 \frac{2^{1-j} \Gamma (j-n) \Gamma (j+n-1)}{\Gamma (j) \Gamma (1-n) \Gamma (n)}
\end{align}
where $(n)_j$ is the Pochhammer symbol.
\par \noindent
Let us list a few examples.
\par \noindent $n = 1$:
\begin{equation}
\partial_{q_i} f(q) = q_i \, f_{(1)} (q) = {q_i \over q} f'(q)
\end{equation}
$n=2$:
\begin{equation}
\partial_{q_i} \partial_{q_j} f(q) = q_i q_j \, f_{(2)} (q) + \delta_{i j} \, f_{(1)}(q) = q_i q_j \left({f''(q) \over q^2} - {f'(q) \over q^3} \right) + \delta_{i j}  {f'(q) \over q}
\end{equation}
$n=3$:
\begin{align}
    \partial_{q_i} \partial_{q_j} \partial_{q_k} f(q) &= q_i q_j q_k f_{(3)}(q) + (q_i \delta_{j k} + q_j \delta_{i k} + q_k \delta_{i j}) f_{(2)}(q) \notag \\
    &= q_i q_j q_k \left( \frac{f^{(3)}(q)}{q^3}-\frac{3 f''(q)}{q^4}+\frac{3 f'(q)}{q^5} \right)\, \notag \\
    &\;\;\;+ (q_i \delta_{j k} + q_j \delta_{i k} + q_k \delta_{i j}) \left({f''(q) \over q^2} - {f'(q) \over q^3} \right) 
\end{align}
$n=4$:
\begin{align}
    \partial_{q_i} \partial_{q_j} &\partial_{q_k} \partial_{q_l}  f(q) = q_i q_j q_k q_l \,f_{(4)}(q) + (\delta_{i k}\delta_{j l} + \delta_{i j} \delta_{k l} + \delta_{j k} \delta_{i l}) \,f_{(2)}(q) \notag \\
    &+ (\delta_{i k} q_j q_l + \delta_{i j} q_k q_l + \delta_{j k} q_i q_l + \delta_{i l} q_j q_k + \delta_{j l} q_i q_k + \delta_{k l} q_i q_j) \,f_{(3)}(q) 
\end{align}
with
\begin{align}
    f_{(4)}(q) &= \frac{f^{(4)}(q)}{q^4}-\frac{6 f^{(3)}(q)}{q^5}+\frac{15 f''(q)}{q^6}-\frac{15 f'(q)}{q^7}, \\
    f_{(3)}(q) &= \frac{f^{(3)}(q)}{q^3}-\frac{3 f''(q)}{q^4}+\frac{3 f'(q)}{q^5}, \\
    f_{(2)}(q) &= \frac{f''(q)}{q^2}-\frac{f'(q)}{q^3}.
\end{align}
\par

\subsection{Scalar contractions with $\vec{a}$, $\vec{k}$ and $\vec{a} \times \vec{k}$ (for Kerr scalar potential)}

We now want to contract the tensorial Fourier transforms mention before with $\vec{a}$, $\vec{k}$ (coming from $\vec{\partial}$ in position space) and $\vec{a} \times \vec{k}$ which are structures appearing on the Kerr potential. We will start with the scalar case where all indices are contracted (i.e. the potential is a scalar function).

The simplest case is the Fourier transform of
\begin{equation}
   V(\vec{x})  = {(a \cdot x)^n \over r^m} = a^{i_1} \cdots a^{i_n} {{x_{i_1} \cdots x_{i_n}} \over r^m}
\end{equation}
will contribute to $V(\vec{p},\vec{k})$ as
\begin{equation}
V(\vec{p},\vec{k}) = (-i)^{n} \, a^{i_1} \cdots a^{i_n} \, \partial_{q_{i_1}} \cdots \partial_{q_{i_n}} \int d^3 x \, {e^{-i \vec{q} \cdot \vec{x}} \over r^m} 
\end{equation}
Since
\begin{equation}
  \partial_{q_{i_1}} \!\cdots \partial_{q_{i_n}} f(q) = \sum_{j=0}^{\lfloor {n \over 2} \rfloor}  \delta_{(i_1 i_2} \cdots \delta_{i_{2j-1} i_{2j}} \, q_{i_{2j + 1}} \cdots \, q_{i_n)} \, f_{(n-j)}(q)
\end{equation}
contracting with $a^{i_1} \cdots a^{i_n}$ gives
\begin{equation}
 a^{i_1} \cdots a^{i_n}  \partial_{q_{i_1}} \!\cdots \partial_{q_{i_n}} f(q) =  \sum_{j=0}^{\lfloor {n \over 2} \rfloor} {n! \over j! \, (n-2j)! \, 2^j }\, (a^2)^j \, (\vec{a} \cdot \vec{q}\,)^{n - 2j}  f_{(n-j)}(q)
\end{equation}
where the symmetry factor is given by the ways of organizing $n$ $a$'s into $j$ $\delta$'s and the remaining number of $q$'s which is $n - 2j$. 
So we have
\begin{align}
    V_{m,n,0,0}(\vec{a},\vec{q},\vec{k}) &\equiv \int e^{i \vec{q} \cdot \vec{x}} {(a \cdot x)^n \over r^m} d^3\vec{x}  \notag \\
    &= (-i)^n \sum_{j=0}^{\lfloor {n \over 2} \rfloor} {n! \over j! \, (n-2j)! \, 2^j }\, (a^2)^j \, (\vec{a} \cdot \vec{q}\,)^{n - 2j}  \,V_{m,n-j}(q)
\end{align}
We may also have one $\vec{k}$ contracted. Namely,
\begin{align}
 k^{i_1} a^{i_2} \cdots a^{i_n}  \partial_{q_{i_1}} \!\cdots \partial_{q_{i_n}} &f(q) =  \sum_{j=0}^{\lfloor {n \over 2} \rfloor} {(n-1)! \over j! \, (n-2j)! \, 2^j } (a^2)^{j-1} \, (\vec{a} \cdot \vec{q}\,)^{n - 2j-1}\,  f_{(n-j)}(q) \notag \\
 &\times \; \Big[ (n - 2j) \, (\vec{k} \cdot \vec{q}) \,  (a^2) \;  + \; (2j) \, (\vec{k} \cdot \vec{a})  \, (\vec{a} \cdot \vec{q}\,) \Big]
\end{align}
for which
\begin{align}
    V_{m,n,1,0}(\vec{a},\vec{q},\vec{k}) &\equiv \int e^{i \vec{q} \cdot \vec{x}} {(\vec{a} \cdot \vec{x})^{n-1} (\vec{x} \cdot \vec{\partial})\over r^m} d^3\vec{x}  \notag \\
    &= (-i)^n i \sum_{j=0}^{\lfloor {n \over 2} \rfloor} {(n-1)! \over j! \, (n-2j)! \, 2^j } (a^2)^{j-1} \, (\vec{a} \cdot \vec{q}\,)^{n - 2j-1}\,  V_{m,n-j}(q) \notag \\
 &\qquad \qquad \times \; \Big[ (n - 2j) \, (\vec{k} \cdot \vec{q}) \,  (a^2) \;  + \; (2j) \, (\vec{k} \cdot \vec{a})  \, (\vec{a} \cdot \vec{q}\,) \Big]
\end{align}
We may also have $\epsilon_{i j k}$ contractions in which case
\begin{align}
    &V_{m,n,0,1}(\vec{a},\vec{q},\vec{k}) \equiv \int e^{i \vec{q} \cdot \vec{x}} {(\vec{a} \cdot \vec{x})^{n-1} (\epsilon^{i j l} x_{i} a_j \partial_l )\over r^m} d^3\vec{x}  \notag \\
    &= (-i)^n i \sum_{j=0}^{\lfloor {n \over 2} \rfloor} {(n-1)! \over j! \, (n-2j-1)! \, 2^j } (a^2)^{j} \, (\vec{a} \cdot \vec{q}\,)^{n - 2j-1}\,  V_{m,n-j}(q) 
  \, (\epsilon^{i j l} q_i a_j k_l) \,
\end{align} 
(the $\epsilon^{i j l} a_i a_j k_l$ piece vanishes).
\par
In the case of contractions with two derivatives $\vec{\partial}$, which is the maximum we can get since the wave equation is second order, we need the result
\begin{align}
 &k^{i_1} k^{i_2} a^{i_3} \cdots a^{i_n}  \partial_{q_{i_1}} \!\cdots \partial_{q_{i_n}} f(q) =  \sum_{j=0}^{\lfloor {n \over 2} \rfloor} {(n-2)! \over j! \, (n-2j)! \, 2^j } (a^2)^{j-2} \, (\vec{a} \cdot \vec{q}\,)^{n - 2j-2}\,  f_{(n-j)}(q) \notag \\
 &\times \; \Big[ (n - 2j)(n-2j-1) \, (\vec{k} \cdot \vec{q})^2 \,  (a^2)^2 \;  + \; 2 (2j) (n - 2j) \,  (\vec{k} \cdot \vec{q})  (\vec{k} \cdot \vec{a})  \, (\vec{a} \cdot \vec{q}\,) \, (a^2) \notag \\
 & \qquad + \; (2j) (2j-2) \, (\vec{k} \cdot \vec{a})^2 (\vec{a} \cdot \vec{q}\,)^2 + (2j) \,(\vec{k})^2\, (\vec{a} \cdot \vec{q}\,)^2 (a^2) \Big]
\end{align}
for which
\begin{align}
&V_{m,n,2,0}(\vec{a},\vec{q},\vec{k}) \equiv \int e^{i \vec{q} \cdot \vec{x}} {(a \cdot x)^{n-2} (x \cdot \partial)^2 \over r^m} d^3\vec{x} \notag \\
&= (-i)^n (i)^2 \sum_{j=0}^{\lfloor {n \over 2} \rfloor} {(n-2)! \over j! \, (n-2j)! \, 2^j } (a^2)^{j-2} \, (\vec{a} \cdot \vec{q}\,)^{n - 2j-2}\,  V_{m,n-j}(q) \notag \\
 &\times \; \Big[ (n - 2j)(n-2j-1) \, (\vec{k} \cdot \vec{q})^2 \,  (a^2)^2 \;  + \; 2 (2j) (n - 2j) \,  (\vec{k} \cdot \vec{q})  (\vec{k} \cdot \vec{a})  \, (\vec{a} \cdot \vec{q}\,) \, (a^2) \notag \\
 &\qquad \qquad + \; (2j) (2j-2) \, (\vec{k} \cdot \vec{a})^2 (\vec{a} \cdot \vec{q}\,)^2 + (2j) \,(\vec{k})^2\, (\vec{a} \cdot \vec{q}\,)^2 (a^2) \Big]
\end{align}
The final possibility is to have one $k$ and one Levi-Civita $\epsilon$,
\begin{align}
&V_{m,n,1,1}(\vec{a},\vec{q},\vec{k}) \equiv \int e^{i \vec{q} \cdot \vec{x}} {(a \cdot x)^{n-2} (x \cdot \partial) (\epsilon^{i j l} x_{i} a_j \partial_l )  \over r^m} d^3\vec{x} \notag \\
&= (-i)^n (i)^2   \sum_{j=0}^{\lfloor {n \over 2} \rfloor} {(n-2)! \over j! \, (n-2j-1)! \, 2^j } (a^2)^{j-1} \, (\vec{a} \cdot \vec{q}\,)^{n - 2j-2}\,  V_{m,n-j}(q)  \notag \\
 &\times \; \Big[ (n-2j-1) \, (\vec{k} \cdot \vec{q}) (\epsilon^{i j l} q_i a_j k_l)  \,  (a^2) \; 
 + \;  (2j)  \,  (\epsilon^{i j l} q_i a_j k_l) (\vec{k} \cdot \vec{a})  \, (\vec{a} \cdot \vec{q}\,) \,  \Big]
\end{align}

(where the derivatives are understood to only act on the wavefunction).
\par

\subsection{Tensorial derivatives of Fourier transform contractions (for photon)}

For tensorial potentials as in the photon example $V_{\mu \nu}$ we may have other possibilities where at most two $x^i$'s are uncontracted in the Fourier transform.
\par
Namely, with one uncontracted index,
\begin{align}
    V^w_{m,n,0,0}(\vec{a},\vec{q},\vec{k}) &\equiv \int x^w e^{i \vec{q} \cdot \vec{x}} {(a \cdot x)^n  \over r^m} d^3\vec{x} = (- i) \, { \partial V_{m,n,0,0} (\vec{a},\vec{q},\vec{k}) \over \partial{q^w}}  \notag \\
    &= (-i)^{n+1} \sum_{j=0}^{\lfloor {n \over 2} \rfloor} {n! \over j! \, (n-2j)! \, 2^j }\, (a^2)^j \, (\vec{a} \cdot \vec{q}\,)^{n - 2j-1} \notag \\
    &\qquad \times \Big[a^w (n- 2j) \, V_{m,n-j}(q) + q^w (\vec{a} \cdot \vec{q}) V_{m,n-j+1}(q) \Big]
\end{align}
with two uncontracted indices,
\begin{align}
    &V^{w z}_{m,n,0,0}(\vec{a},\vec{q},\vec{k}) \equiv \int x^w  x^z e^{i \vec{q} \cdot \vec{x}} {(a \cdot x)^n  \over r^m} d^3\vec{x} = (- i)^2 \, { \partial^2 V_{m,n,0,0} (\vec{a},\vec{q},\vec{k}) \over \partial{q^w} \partial{q^z}}  \notag \\
    &= (-i)^{n+2} \sum_{j=0}^{\lfloor {n \over 2} \rfloor} {n! \over j! \, (n-2j)! \, 2^j }\, (a^2)^j \, (\vec{a} \cdot \vec{q}\,)^{n - 2j-2} \notag \\
    &\times \Big[a^w a^z (n- 2j) (n-2j-1) \, V_{m,n-j}(q) +
   (a^w q^z + a^z q^w) (n-2j) (\vec{a} \cdot \vec{q}\,) V_{m,n-j+1}(q) \notag \\
    &\qquad + \delta^{w z} (\vec{a} \cdot \vec{q}\,)^2  V_{m,n-j+1}(q) \;+ \; q^w q^z (\vec{a} \cdot \vec{q}\,)^2 V_{m,n-j+2}(q) \Big]
\end{align}

\bibliographystyle{JHEP}
\bibliography{Born}

\end{document}